 \documentclass{IEEEtran}
 \IEEEoverridecommandlockouts
 \usepackage{cite}
 \usepackage{amsmath,amssymb,amsfonts}
 \usepackage{algorithmic}
 \usepackage{graphicx}
 \usepackage{textcomp}
 \usepackage{xcolor}
 \usepackage{setspace}
 \usepackage{enumitem}
 \newcommand*{\J}{\jmath}%
 \newcommand{\diff}{\mathop{}\!d}
 \def\BibTeX{{\rm B\kern-.05em{\sc i\kern-.025em b}\kern-.08em
 		T\kern-.1667em\lower.7ex\hbox{E}\kern-.125emX}}
 \usepackage{subfigure}
 \newtheorem{my_theorem}{Theorem}
 
 \newtheorem{my_lemma}{Lemma}

 \title{Unified Performance Assessment of Optical Wireless  Communication over Multi-Layer Underwater Channels}
 \author{Ziyaur Rahman,~\IEEEmembership{Gradute Student Member,~IEEE}, Neel Vipulbhai Tailor, S.~M.~ Zafaruddin,~\IEEEmembership{Senior Member,~IEEE,} and
 	V.~K.~ Chaubey,~\IEEEmembership{Senior Member,~IEEE}
 	\thanks{ A part of this paper on the generalized gamma distribution for the multi-layer oceanic turbulence  is under review  in the 2022 IEEE 95th Vehicular Technology Conference: VTC2022-Spring to be held in Helsinki, Finland 19-22 June 2022 \cite{Suhrid2022_arxiv}.}	
 	\thanks{This work was supported in part by the 
 		Science and Engineering Research Board (SERB), India under MATRICS Grant MTR/2021/000890 and Start-up Research Grant SRG/2019/002345.
 		
 		The authors	are  with  the Department of Electrical and Electronics Engineering, Birla Institute of Technology and Science, Pilani, Pilani-333031, Rajasthan, India. (Email: p20170416@pilani.bits-pilani.ac.in, f20190152@pilani.bits-pilani.ac.in, syed.zafaruddin@pilani.bits-pilani.ac.in, vkc@pilani.bits-pilani.ac.in .}
 	
 	\thanks{}}
  
\begin{document}
 	\maketitle
   \begin{abstract}
    In this paper, we  model the multi-layer vertical underwater link as a  cascaded channel and unify the performance analysis for the underwater optical communication (UWOC) system using   generalized Gamma (GG), exponential GG (EGG), exponentiated Weibull (EW), and Gamma-Gamma ($\Gamma\Gamma$) oceanic turbulence models.   We derive  unified analytical expressions for probability density function (PDF) and cumulative distribution function (CDF) for the signal-to-noise ratios (SNR) considering independent and non-identical (i.ni.d.) turbulent models and zero bore-sight model for pointing errors. We develop  performance metrics of the considered UWOC system using outage probability, average bit error rate (BER), and ergodic capacity with  asymptotic expressions for outage probability and average BER. We develop  the diversity order of the proposed system to provide  a better insight into the system performance at a high SNR. We also integrate a terrestrial OWC (TOWC) subjected to the combined effect of  generalized Mal\'aga atmospheric turbulence, fog-induced random path gain, and pointing errors to communicate with the UWOC link using the fixed-gain amplify-and-forward (AF) relaying. We analyze the performance of the mixed TWOC and multi-layer UWOC system by deriving PDF, CDF, outage probability, and average BER using the bivariate Fox H-function.  We use Monte-Carlo simulation results to validate our exact and asymptotic expressions and demonstrate the performance of the considered underwater UWOC system  using measurement-based parametric data available for   turbulent oceanic channels.	
   	
   \end{abstract}
   
   \begin{IEEEkeywords}	
   	Cascaded channels,  multi-layer channels, Mellin's transform,  performance analysis, oceanic turbulence,  UWOC, vertical link.
   \end{IEEEkeywords}	
   
   
   \section{Introduction}
   Underwater optical communication (UWOC) is a potential solution for broadband connectivity in oceans and seas for underwater applications  providing high data rate transmission with low latency and high reliability \cite{Gussen2016, Kaushal2016, Zhaoquan2017}.  It is a promising technology for  underwater data transmission  providing higher throughput with low latency and high reliability than radio frequency (RF) and acoustic wave communication systems. The underwater optical communication (UWOC) system transmits  data in an  unguided water environment using the wireless optical carrier   for military, economic and scientific applications \cite{Zhaoquan2017}. Despite several advantages of the UWOC,  the underwater link suffers from  signal attenuation, oceanic turbulence, and pointing errors.  The signal attenuation occurs due to the  molecular absorption and scattering effect of each photon propagating through water, generally modeled by the extinction coefficient. Oceanic turbulence is the effect of random variations in the refractive index of the  UWOC channel caused by  random variations of water temperature, salinity, and air bubbles.  Pointing errors can also be detrimental to UWOC transmissions  due to misalignment between the transmitter and detector apertures. Therefore, it is  desirable to analyze the UWOC systems over various underwater channel impairments  for an effective system design.

   As is for any communication system, recent works  developed theoretical and experimental characterization of turbulence-induced fading under various underwater conditions \cite{Tang2013, Yi2015, Nabavi2018, Farwell2012, Vahid2018,  Oubei2017, Zedini2019}.  Research outcomes in \cite{Tang2013, Yi2015,Nabavi2018} demonstrate that the log-normal distribution efficiently models weak oceanic turbulence similar to the modeling of weak atmospheric turbulence for terrestrial OWC links.  The authors in \cite{Farwell2012} demonstrated higher oceanic turbulence since the scintillation index for an optical wave is very high over several meters of underwater propagation. In \cite{Vahid2018}, the authors presented a holistic experimental view on the statistical characterization of oceanic turbulence in UWOC systems, considering the effect of the temperature gradient, salinity, and air bubbles. They used various statistical distributions such as log-normal, Gamma, Weibull, Exponentiated Weibull (EW), Gamma-Gamma ($\Gamma \Gamma$), and generalized Gamma (GG)  to model  underwater turbulence channels. Further, experimental investigations projected  the  GG distribution and EW as  more generic  models and valid for various  underwater channel conditions  \cite{Vahid2018,Oubei2017}. Recently, \cite{Zedini2019} used experimental data to propose
  the mixture exponential-generalized Gamma (EGG) distribution for oceanic turbulence caused by air bubbles and   temperature gradient for UWOC channels, which perfectly matches the measured   data, collected under different channel conditions ranging from   weak to strong turbulence conditions.

   There has been tremendous research on the performance assessment of UWOC systems \cite{Xiaobin2020,Gercekcioglu2014,Vahid2016,Mingjian2016,Vahid2015,Vahid2017, Jamali2017,Azadeh2017,Ahmadirad2018,Zedini2019,Zedini2020_TCOM} and the mixed system consisting of UWOC and terrestrial networks    \cite{Lei2020, Anees2019, Li2020,Li2021, Ansari2019,Yang2021}.  
   In those mentioned above and related research, a single layer of oceanic turbulence channel over the  entire transmission range has been considered. However, experimental results reveal ocean stratification, i.e., the temperature gradient and salinity are depth-dependent (typically varying between a few meters to tens of meters), resulting in  many non-mixing  layers with different oceanic turbulence \cite{Elamassie2019}.  Thus, considering multiple oceanic layers for vertical transmissions may  provide a more realistic performance assessment for UWOC systems. In \cite{Elamassie2018,Mohammed2018,Elamassie2019,Elamassie2019pe},  the author analyzed the performance vertical UWOC links by cascading the end-to-end link as the concatenation of multiple layers considering both log-normal and $\Gamma\Gamma$ oceanic turbulent channels for each layer. They used the method of induction to analyze the cascaded $\Gamma\Gamma$  channel, which may not be readily applicable to other  channel models.  To the best of the authors’ knowledge,  no analyses available for the outage probability, average BER, and ergodic capacity of a multi-layer UWOC system considering GG, EGG, and EW oceanic turbulence models. Further, it is desirable to consider a more generalized model for the  terrestrial OWC (TOWC) that includes the combined effect of Mal\'aga atmospheric turbulence, fog-induced random path gain, and pointing errors to study the mixed UWOC-TWOC transmission. It should be mentioned that   the terrestrial  OWC link might be affected by  foggy conditions near the ocean/sea, and consideration of deterministic path loss may underestimate/overestimate the performance of the considered system \cite{Rahman2021TVT}.

   This paper presents a unified performance analysis of a vertical UWOC system under the combined effect of multilayer underwater turbulence channels and pointing errors.  The major contributions of the proposed work are summarized as follows:
   \begin{itemize}
   	\item   We apply the Mellin inverse transform to develop   analytical expressions for probability density function (PDF) and cumulative distribution function (CDF) for the signal-to-noise    ratios (SNR) of UWOC system unifying  GG, EGG, EW, and $\Gamma\Gamma$ oceanic turbulent models and zero bore-sight model for pointing errors.
   	
   	\item We use the derived statistical results to develop  performance metrics of the considered UWOC system using outage probability, average bit error rate (BER), and ergodic capacity with  asymptotic expressions for the outage probability and average BER to determine the diversity order of the proposed system for a better insight into the system performance.
   	\item  We integrate a generalized terrestrial OWC (TOWC) link subjected to the combined effect of  generalized Mal\'aga atmospheric turbulence, fog-induced random path gain, and pointing errors to communicate with the UWOC link using the fixed-gain amplify-and-forward (AF) relaying. We analyze the performance of the mixed TWOC-UWOC by deriving PDF, CDF, outage probability, and average BER using bivariate Fox H-function. 
   	\item We use numerical and simulation analysis to validate our derived expressions and demonstrate the performance of the considered UWOC system for various parameters of interest.
   \end{itemize}
\subsection{Related Work}
There is rich literature considering the performance analysis for the single UWOC link under various oceanic turbulence conditions \cite{ Zedini2019, Xiaobin2020,Gercekcioglu2014,Vahid2016,Mingjian2016,Vahid2015,Vahid2017,Jamali2017, Azadeh2017,Ahmadirad2018}.  In \cite{Zedini2019}, the authors analyzed  the outage probability, average BER, and ergodic capacity for UWOC by modeling the underwater optical turbulence channel  using the EGG distribution. The  authors in \cite{Xiaobin2020}  provided an overview of  various challenges associated with UWOC and proposed positioning, acquisition, and tracking scheme to mitigate the effect of pointing errors under turbulent channels. The average bit-error-rate (BER) performance under weak log-normal distributed turbulence channels was  presented in  \cite{Gercekcioglu2014}. The authors in \cite{Vahid2016} characterized a relay-assisted UWOC with optical code division multiple access (OCDMA) over log-normal turbulent channels. An analytic expression for the channel capacity of an orbital angular momentum (OAM) based free-space optical (FSO) communication in weak oceanic turbulence was developed in  \cite{Mingjian2016}.   The authors in \cite{Vahid2015,Vahid2017}  analyzed the performance of  multi-input and multi-output (MIMO) UWOC systems over  log-normal turbulent channels.   Further, a multihop UWOC system was investigated in \cite{Jamali2017}. The outage probability of a multiple decode-and-forward (DF) relay-assisted  UWOC system with an on-off keying (OOK) modulation  was studied in \cite{Azadeh2017}.   The various optical turbulence models like log-normal, Gamma, $K$, Weibull, and exponentiated Weibull	distributions have been used to analyze the performance of underwater wireless optical communication (UWOC) systems \cite{Ahmadirad2018}. It should be emphasized that the related work on the UWOC system consider a single channel and that there is limited research on the vertical cascaded  using log-normal and $\Gamma \Gamma$ turbulence models \cite{Elamassie2018,Mohammed2018,Elamassie2019,Elamassie2019pe}.

 Since the advent of UWOC, there is an increased interest to offload the underwater data to a terrestrial network using RF technology \cite{Lei2020, Anees2019,Li2020, Li2021,Ansari2021_TVT} and terrestrial OWC \cite{Ansari2019, Yang2021}.  The authors in \cite{Lei2020, Anees2019} considered Nakagami-$m$ fading for the radio frequency (RF) and   EGG turbulence for the UWOC and analyzed the outage probability and average BER for the mixed RF-UWOC system.  In \cite{Li2020}, the fixed-gain AF relaying was used to mix the RF link over generalized-K distributed fading and the EGG distributed UWOC link. In \cite{Li2021}, the authors analyzed the performance of dual-hop RF-UWOC system assisted by an unmanned aerial vehicle (UAV) using both fixed-gain AF relaying and decode-and-forward (DF) relaying schemes. The use of multiple input multiple output (MIMO) for RF transmission under mixed RF-UWOC was studied in \cite{Ansari2021_TVT}. Terrestrial transmission using optical wireless was  recently investigated  in \cite{Ansari2019, Yang2021}.  The authors in \cite{Ansari2019} studied the  outage probability of a mixed terrestrial OWC link with multi-sensor UWOC considering  weak 	oceanic and atmospheric turbulence conditions. Recently, the authors in \cite{Yang2021}  used the fixed-gain AF relaying to   mix TOWC-UWOC communication system  by modeling  the TOWC channel using  Gamma-Gamma atmospheric turbulence with pointing errors and the EGG distributed  UWOC channel with pointing errors.    

\subsection{Notations and Organization of the Paper} 
  
 \begin{table}[t]
  	\caption{The List of Main  Notations}
 	\label{table:notation_parameters}
 	\centering
 	\begin{tabular}{|p{3.1cm}|p{4.5cm}|}
 		\hline
 		${(\cdot)}_{T}$ & Notation for the TWOC link \\ \hline  ${(\cdot)}_{U}$ & notation for the UWOC link\\ 
 		\hline
 		${[\cdot]}_{i}$ & Notation for the $i$-th element \\ \hline
 	
 		$l_T$ &  TWOC link distance \\ \hline 
 		$l_U$ &  UWOC link distance \\ \hline 
 	 		 $\bar{\gamma}$ &  Average SNR\\\hline
 		 $\gamma$  & Instantaneous SNR\\ \hline 
 		 $k$, 	$\beta^f$ &  Parameters for foggy channel\\
 		\hline
 		 		 $\alpha^{{\scriptscriptstyle  M}}$, $\beta^{{\scriptscriptstyle  M}}$,  $A^{m\rm g}$, $b_m^M$, $g^M$, $\Omega^M $ & Mal\'aga distribution parameters \\ \hline 	
 		$\rho$, $A$ & Pointing errors parameters \\ \hline 	
 		
 		$\omega$, $\lambda$, $a$, $d$, $p$ & EGG and GG distribution parameters \\ \hline 
 		$\alpha^E$, $\beta^E$, $\eta^E$ & EW distribution parameters \\ \hline 
 		$\alpha^G$, $\beta^G$ & $\Gamma \Gamma$ distribution parameters \\ \hline 
 	 		$\mathbb{E}[\cdot]$ & Expectation operator \\ \hline 
 		$\Gamma(a)$ & $\int\limits_{0}^{\infty}t^{a-1} e^{-t}dt$ \\ \hline 		$G_{p,q}^{m,n}\left(\begin{matrix}(a_k)^p_{k=1}  \\ (b_k)^q_{k=1}\end{matrix}   \middle \vert z \right)$ & Meijer's G-function \\ \hline$H_{p,q}^{m,n} \left[ \begin{matrix}(a_k,A_k)^p_{k=1}  \\ (b_k,B_k)^q_{k=1}\end{matrix}  \middle \vert z \right]$ & Fox's H-function \\ \hline 
 		
 	\end{tabular}	
 \end{table}

 We list the main notations  in Table \ref{table:notation_parameters}.
 
  The paper is organized as follows. Section II discusses the channel models for both terrestrial and underwater optical communications. In Section III presents statistical results for the multi-layer UWOC system. In Section IV, the performance of the mixed TOWC-UWOC system in terms of outage probability and average BER is  analyzed. In Section V, we present the numerical and simulation analysis of the proposed system. Finally, important conclusions are stated in Section VI.

   \section{System Model}
    \begin{figure}[t]	
   	\centering
   	\includegraphics[scale=0.35]{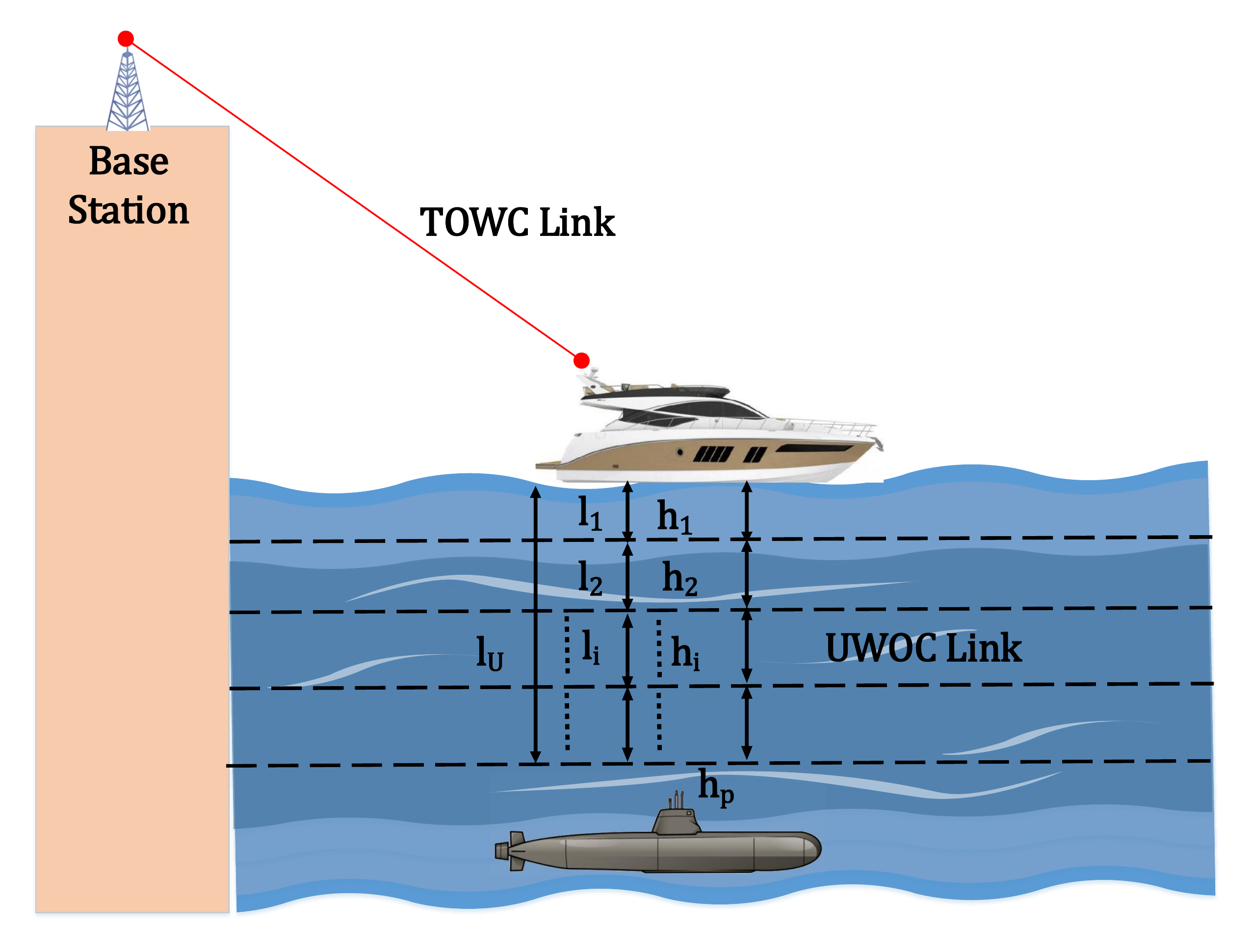}
      	\caption{ Mixed TWOC-UWOC system with multi-layer UWOC channel.}
   	\label{system_model}	
   \end{figure}
   We consider a mixed terrestrial and underwater optical communication system integrated through a fixed-gain AF relaying protocol,   as shown in Fig~\ref{system_model}. Assume that a source on the land intends to communicate a signal $s$ with an underwater submarine.   We use the non-coherent intensity modulation/direct detection (IM/DD) scheme, where the photodetector detects changes in the light intensity without employing a local oscillator. It is known that the  heterodyne detection (HD) requires complex processing of  mixing  the received signal with a coherent signal produced by the local oscillator \cite{Melchior1970}. In the following two subsections, we describe channel and system models for terrestrial and  underwater OWC systems.
   \subsection{Terrestrial OWC}
 In the first hop, we assume that the transmitted signal undergoes three types of fading: atmospheric turbulence-induced, pointing errors, and random fog. Thus, the received signal $y_T$ at the relay  is given by
  \begin{eqnarray}
 	y=h_Ts+ w_T
 	\label{eq:sys_T}
 \end{eqnarray}
 where $h_T$ is the channel coefficient (including fog-induced path gain, atmospheric turbulence, and pointing errors) for the  terrestrial link and $w_T$ is the additive white gaussian noise (AWGN) with variance $\sigma_{w_T}^2$
 Assuming  generalized Mal\'aga distribution for atmospheric turbulence, fog-induced random path loss, and zero bore-sight pointing errors, then the PDF of the SNR $\gamma_{T}$for the terrestrial link is given by    \cite{Chapala2021}
   \begin{eqnarray}
   	&f_{\gamma_{T}}(x)=\frac{z^k\rho_T^2A^{\rm{mg}}}{4 \gamma}  \sum_{m_1=1}^{\beta^M}b^M_{m_1}\nonumber \\&G_{1+k,3+k}^{3+k,0} \left(\begin{array}{c} \rho_T^2+1, \{z+1\}_1^k\\ \rho_T^2, \alpha^M, m_1, \{z\}_1^k\end{array} \left| \frac{\alpha^M\beta^M \sqrt{\gamma}}{(g^M\beta^M+\Omega^M)A_T\sqrt{\bar{\gamma}_T}}\right.\right)
   	\label{eq:pdf_fpt_twoc}
   \end{eqnarray}	
      where $\bar{\gamma}_T$ is the average SNR,  $\{\alpha^{{\scriptscriptstyle  M}}, \beta^{{\scriptscriptstyle  M}},  A^{m\rm g}, b_{m_1}^M, g^M, \Omega^M \}$ are Mal\'aga parameters \cite{malaga2011},  $\{\rho_T, A_T\}$ are pointing error parameters \cite{Farid2007}, and $\{z, k\}$ specifies the effect of fog on the signal transmission \cite{Esmail2017_Access}.    
   \subsection{Underwater OWC}
   In the second hop, we employ a fixed-gain AF relay with gain parameter $G_R$ to forward the received to the destination over underwater channel.  The gain selection  $G_R$ can be entirely blind for a duration or using a semi-blind approach where it can be obtained using statistics of received signal power of the first hop (i.e., TOWC link). We consider the UWOC system by splitting  the entire transmission channel in $N$ distinct layers in succession, resulting in $N$ vertical links,  as depicted in Fig~\ref{system_model}.  Thus, the received electrical signal $y_U$ at the destination can be expressed as:
   \begin{eqnarray}
   	y_U=h_lh_{p}\left[\prod_{i=1}^{N}h_{i}\right]G_Ry_T+ w_U
   	\label{eq: System Model main eq}
   \end{eqnarray}
   where $h_l=e^{-\alpha l}$ is the  path gain with link distance $l$ (in \rm{m}) and extinction attenuation coefficient $\alpha$,  $h_{p}$ models  pointing errors, $h_c=\prod_{i=1}^{N}h_{i}$ is the cascaded channel with $h_i$ as the $i$-th layer of vertical link, and $w_U$ is the  AWGN  with variance $\sigma_{w_U}^2$.

   The PDF of zero-boresight pointing errors fading $h_{p}$ is given as \cite{Farid2007}:
   \begin{eqnarray}
   	f_{h_{p}}(x) = \frac{\rho^2}{A_{0}^{\rho^2}}h_{p}^{\rho^{2}-1},0 \leq h_{p} \leq A_0
   	\label{eq:pdf_hp}
   \end{eqnarray}
   where $A_0=\mbox{erf}(\upsilon)^2$ with $\upsilon=\sqrt{\pi/2}\ r/\omega_{z}$, $r$ is the aperture radius and $\omega_{z}$ is the beam width, and $\rho = {\frac{\omega_{{z}_{\rm eq}}}{2 \sigma_{s}}}$ with  $\omega_{{z}_{\rm eq}}$ as the equivalent beam width at the receiver and $\sigma^2_{s}$ as the variance of pointing errors displacement characterized by the horizontal sway and elevation \cite{Farid2007}.

 Fading coefficients $h_i$, $i=1,2, \cdots, N$ associated with  different layers are modeled using various statistical distributions such as generalized Gamma, EGG, GG, and   EW among others for different underwater conditions \cite{Vahid2018,Zedini2019}. 
 
 The PDF of the channel coefficient for the generalized Gamma is given as   
   \begin{flalign}
   	f_{h_{i}}(x)=\frac{p_{i}}{{a_{i}}^{d_{i}}\Gamma\Big(\frac{d_{i}}{p_{i}}\Big)}{x}^{d_{i}-1}\exp\bigg(-\Big(\frac{x}{a_{i}}\Big)^{p_{i}}\bigg)
   	\label{eq:h_gg}
   \end{flalign}
   where $a_{i}$, $d_{i}$, and $p_{i}$ are distribution parameters for the $i$-th layer to model different oceanic turbulence scenarios, as given in \cite{Vahid2018} (see Table-I, Table-II, and Table-III). As such, $p_i=1$ in 	\eqref{eq:h_gg} denotes a Gamma distribution representing  a thermally uniform UWOC channel. 
   
 Recently, \cite{Zedini2019} proposed   EGG distribution (i.e., the combined exponential and generalized Gamma) for the oceanic turbulence with the PDF:
\begin{eqnarray}
	&f_{h_{i}}(x) = \frac{\omega_i}{\lambda_i}\exp(-\frac{x}{\lambda_i})+\nonumber \\& (1-\omega_i)\frac{p_{i}}{{a_{i}}^{d_{i}}\Gamma\Big(\frac{d_{i}}{p_{i}}\Big)}{x}^{d_{i}-1}\exp\bigg(-\Big(\frac{x}{a_{i}}\Big)^{p_{i}}\bigg)
	\label{pdf_uw}
\end{eqnarray}
where $\omega_i$ is the mixture
coefficient of the distributions (i.e, $0<\omega_i>1$), $\lambda_i$ is the exponential distribution parameter. Note that experimental data is available for $a_i, p_i$, and $d_i/p_i$.   

  Further, the PDF of the channel coefficient using the three-parameter EW distribution to model the oceanic turbulence is given by:
   \begin{eqnarray}
   	&f_{h_i}(x)=\frac{\alpha_i^E\beta_i^E}{\eta_i^E}\left( \frac{x}{\eta_i^E}\right)^{({\beta_i^E}-1)}\exp\left[-\left(\frac{x}{\eta_i^E}\right)\right]^{\beta_i^E}\nonumber\\&\left[1-\exp\left[-\left(\frac{x}{\eta_i^E}\right)^{\beta_i^E}\right]\right]^{({\alpha_i^E}-1)}
   	\label{eq:pdf_ew}
   \end{eqnarray}
where $\beta_i^{E} > 0$ denotes the shape parameter of the scintillation index (SI), $\eta_i^E > 0$ is a scale parameter, and $\alpha_i^E > 0$ is an extra shape parameter  dependent on the receiver aperture size \cite{Barrios2012}. 
    
Experimental studies also found  the  Gamma-Gamma PDF \cite{Andrews20015} a good fit for oceanic turbulent channel \cite{Vahid2018}:
   \begin{flalign}
   	&f_{h_i}(x)=\frac{2(\alpha_i^{G}\beta_i^G)^{(\alpha_i^G+\beta_i^G)/2}}{\Gamma({\alpha_i^G})\Gamma({\beta_i^G})}x^{\frac{\alpha_i^G+\beta_i^G}{2}-1}\nonumber\\&K_{\alpha_i^G-\beta_i^G}(2(\alpha_i^G\beta_i^G x)^{1/2})
   	\label{eq:pdf_gamma_gamma}
   \end{flalign}
In what follows, we unify the performance of multi-layer UWOC system over different oceanic turbulence models.  
   \section{Multi-Layer UWOC }
   In this section, we develop   statistical analysis for the multi-layer underwater turbulence channel unifying generalized Gamma, EGG, GG, and EW oceanic turbulence  models. We also analyze the UWOC performance by deriving  outage probability, average BER, and ergodic capacity  demonstrating the impact of multi-layer modeling  for the vertical channel.
   \subsection{Statistical Results}
   	  We use the inverse Mellin transform to find the PDF of $h_c=\prod_{i=1}^{N}h_{i}$ for a given oceanic turbulence. If $\mathbb{E}[X^n]$ denotes the $n$-th moment, where $\mathbb{E}[\cdot]$ denotes the expectation operator, then the inverse Mellin transform results the PDF of a random variable $X$ as
   	 \begin{equation}
   	 	f_X(x)=\frac{1}{2\pi ix}\int_{\zeta-i\infty}^{\zeta+i\infty}x^{-n}\mathbb{E}[X^n]dn
   	 	\label{eq:Nth_order_with_Inverse_Mellin}
   	 \end{equation}
   	 where $\zeta-i\infty$ to $\zeta+i\infty$ denotes the line integral. It should be mentioned that Mellin transform has been used to analyze the product of $N$ random variable for different applications \cite{Kaddoum2018, Chapala2021,Chapala2021_LCOMM,bhardwaj2021performance}. Similar to \cite{Vahid2018}, we assume that turbulence channels are independent for each layer to get the $n$-th order moment for $h_c$  as	
   	 \begin{eqnarray}
   	 	&\mathbb{E}[h_c^n]=\prod_{i=1}^{N}\int_{0}^{\infty}h_i^n f_{h_i}(h_i)dh_i
   	 	\label{eq:nth_order_moment_of_hi}
   	 \end{eqnarray}
    In the following theorem, we use 	\eqref{eq:nth_order_moment_of_hi} in 	\eqref{eq:Nth_order_with_Inverse_Mellin} to derive the PDF for various oceanic turbulence. 
 \begin{my_theorem}
	An unified expression for the PDF of multi-layer UWOC channel distributing according to GG, EGG,  $\Gamma \Gamma$, and EW  is given by
\begin{flalign}
f_{h_c}(h_c)=&\prod_{i=1}^{N}\mathcal{A}_i \sum_{j=0}^{\mathcal{P}}\mathcal{B}_{i,j}\nonumber\\&\frac{1}{h_c}H_{0,q}^{m,0} \left[\begin{matrix} - \\\{\mathcal{D}_{i_1,j}\}_{i_1=1}^N \end{matrix} \bigg|\prod_{i_2=1}^{N} \mathcal{E}_{i_2,j} h_c\right]
\label{eq:unified_uw_pdf}
\end{flalign}
where  $\mathcal{A}_i$, $\mathcal{P}$, $\mathcal{D}_{i_1,j}$, $\mathcal{E}_{i_2,j}$, $m$, and $q$ are given in Table \ref{table1}.
\begin{table}[t]
	\caption{Unified Parameters}
	\centering
		\label{table1}
	\begin{tabular}{|c|p{5.5cm}|}
		\hline
		Oceanic Models & Unified Parameter Description\\
		\hline
		\hline
			EGG &
	 $\mathcal{P}=1, 	\mathcal{A}_i= 1, \mathcal{B}_{i,0} =\omega_i$, $\mathcal{B}_{i,1} =\frac{(1-\omega_i)}{\Gamma\left(\frac{d_i}{p_i}\right)}$, $m=N$, $q=N$, $\mathcal{D}_{i_1,0}=(1,1)$, $\mathcal{D}_{i_1,1}=\left(\frac{d_{i_1}}{p_{i_1}},\frac{1}{p_{i_1}} \right)$, $\mathcal{E}_{i_2,0}=\frac{1}{\lambda_{i_2}}$, $\mathcal{E}_{i_2,1}=\frac{1}{a_{i_2}}$\\
		\hline
			GG & 
	 $\mathcal{P}=0, 	\mathcal{A}_i= \frac{1}{\Gamma\left(\frac{d_i}{p_i}\right)},\mathcal{B}_{i,j} =1$, $m=N$, $q=N$, $\mathcal{D}_{i_1,j}=\left(\frac{d_{i_1}}{p_{i_1}},\frac{1}{p_{i_1}} \right)$, $\mathcal{E}_{i_2,j}=\frac{1}{a_{i_2}}$\\
		\hline
		EW  & $\mathcal{P}=\infty, \mathcal{A}_i=\Gamma(\alpha^E_i+1),  \mathcal{B}_{i,j} =\frac{(-1)^j}{(j+1)!\Gamma(\alpha^E_i-j)}$, $m=N$, $q=N$, $\mathcal{D}_{i_1,j}=\left(1,\frac{1}{\beta^E_{i_1}}\right)$, $\mathcal{E}_{i_2,j}=\frac{(j+1)^{\frac{1}{\beta^E_{i_2}}}}{\eta^E_{i_2}}$\\
	
		\hline
		$\Gamma\Gamma$ &
	 $\mathcal{P}=0,\mathcal{A}_i= \frac{1}{\Gamma(\alpha^G_i)\Gamma(\beta^G_i)}, \mathcal{B}_{i,j} =1$, $m=2N$, $q=2N$, $\mathcal{D}_{i_1,j}=\{(\alpha^G_{i_1},1),(\beta^G_{i_1},1)\}$, $\mathcal{E}_{i_2,j}=\alpha^G_{i_2}\beta^G_{i_2}$\\
		
		\hline	
		\hline
	\end{tabular}	
\end{table}
\end{my_theorem}

\begin{IEEEproof}
See Appendix A.

\end{IEEEproof}  
The derived PDF  in Theorem 1  is represented using  a single variate Fox-H function, which can be computed efficiently through computational software.

 Next, we use the statistical result of Theorem 1 to analyze the multi-layer UWOC performance. Assuming IM/DD technique and on-off keying (OOK) modulation with $x\in \{0,\sqrt{2}P_t\}$  and $P_t$ as average transmitted optical, the instantaneous  received electrical SNR is given by \cite{Farid2007}
\begin{eqnarray}
	\gamma_U=\frac{P^2_t h^2_l h^2}{\sigma_{w}^2}=\bar{\gamma}_U h^2
	\label{eq: inst_snr}
\end{eqnarray} 
where $h=h_ch_p$ is the combined channel and $\bar{\gamma}_U=\frac{P^2_t h^2_{l_U}}{\sigma_{wT}^2}$ is the average electrical SNR. Note that  $P_t^2$ in \eqref{eq: inst_snr} is attributed to the detection type IM/DD and becomes $\gamma_U=\frac{P_t h_{l_U} h}{\sigma_{wT}^2}$ for the HD technique \cite{Melchior1970, Chapala2021}.
 
In the following Lemma, we provide PDF and CDF of the SNR for the UWOC vertical links under the combined effect of the oceanic turbulence and pointing errors:
\begin{my_lemma}\label{theorem_unified_hp}
	Unified  expressions for PDF and CDF of the SNR for multi-layer UWOC system with pointing errors are given as:
	\begin{eqnarray}
		&f_{\gamma_{U}}(\gamma)=\prod_{i=1}^{N}\frac{\rho_U^2\mathcal{A}_i}{2} \sum_{j=0}^{\mathcal{P}}\mathcal{B}_{i,j}\nonumber\\&\frac{1}{\gamma}H_{1,q+1}^{m+1,0} \left[\begin{matrix} (1+\rho_U^2) \\\{\mathcal{D}_{i_1,j}\}_{i_1=1}^N, (\rho_U^2,1) \end{matrix} \bigg|\prod_{i_2=1}^{N} \frac{\mathcal{E}_{i_2,j}}{A_U} \sqrt{\frac{\gamma}{\bar{\gamma}_U}}\right]
		\label{eq:pdf_unified_hp}
	\end{eqnarray}
	\begin{eqnarray}
		&F_{\gamma_{U}}(\gamma)=\prod_{i=1}^{N}\frac{\rho_U^2\mathcal{A}_i}{2} \sum_{j=0}^{\mathcal{P}}\mathcal{B}_{i,j}\nonumber\\&H_{2,q+2}^{m+1,1} \left[\begin{matrix} (1,1),(1+\rho_U^2) \\\{\mathcal{D}_{i_1,j}\}_{i_1=1}^N, (\rho_U^2,1), (0,1) \end{matrix} \bigg|\prod_{i_2=1}^{N} \frac{\mathcal{E}_{i_2,j}}{A_U} \sqrt{\frac{\gamma}{\bar{\gamma}_U}}\right]
		\label{eq:cdf_unified_hp}
	\end{eqnarray}
\end{my_lemma}

\begin{IEEEproof}	
	See Appendix B.
\end{IEEEproof}

\subsection{Outage Probability}
Outage probability is a performance metric which demonstrate the effect of fading channel on the communication systems. 
It is defined as the probability that the SNR falls below a certain threshold $\gamma_{\rm th}$ and is given as
\begin{eqnarray}
P_{\text{out}}=P(\gamma<\gamma_{\text{th}})=F_{\gamma}(\gamma_{\text{th}})
\label{eq:outage_prob_general}
\end{eqnarray}
where $\gamma_{\rm th}$ is the SNR threshold. Substituting \eqref{eq:cdf_unified_hp} in \eqref{eq:outage_prob_general} yields an exact expression for the outage probability. The asymptotic expression for the outage probability in the high SNR regime $\bar{\gamma}_U\to \infty$ can be derived by applying  \cite[eq. $1.8.4$]{Kilbas}:
\begin{eqnarray}
&P_{\rm{out}}^{\infty}=\prod_{i=1}^{N}\frac{\rho_U^2\mathcal{A}_i}{2} \sum_{j=0}^{\mathcal{P}}\mathcal{B}_{i,j}\sum_{k=1}^{m+1}\nonumber\\&\frac{\left( \frac{\mathcal{E}_{i_2,j}}{A_U} \sqrt{\frac{\gamma_{\rm th}}{\bar{\gamma}_U}}\right)^{\frac{b_k}{\beta_k}}\prod_{j=1, j\neq k}^{m+1}\Gamma\left(b_j-b_k\frac{\beta_j}{\beta_k}\right)\Gamma\left(\frac{b_k}{\beta_k}\right)}{\beta_k\Gamma\left(1+\rho_U^2-\frac{b_k}{\beta_k}\right)\Gamma\left(1+\frac{b_k}{\beta_k}\right)}
\label{eq:outage_prob_asymp}
\end{eqnarray}
where $b_j=b_k=\{\rho_U^2, [\mathcal{D}_{i_1,j}]_1\}$ and $\beta_j=\beta_k=\{1, [\mathcal{D}_{i_1,j}]_2\}$.

The dominant SNR terms of \eqref{eq:outage_prob_asymp} provides the diversity of proposed system as ${DO}_{{\rm out}}=\min\{\sum_{i=1}^{N}\frac{b_k}{2\beta_k}\}$. Using Table \ref{table1}, the diversity order for the EGG, GG, EW and $\Gamma\Gamma$ oceanic turbulence  can be derived as: ${DO}_{{\rm EGG}}=\min\{\frac{N}{2},\sum_{i=1}^{N}\frac{d_i}{2}, \frac{\rho_U^2}{2}\}$, ${DO}_{{\rm GG}}=\min\{\sum_{i=1}^{N}\frac{d_i}{2}, \frac{\rho_U^2}{2}\}$, ${DO}_{{\rm EW}}=\min\{\sum_{i=1}^{N}\frac{\beta_i^E}{2}, \frac{\rho_U^2}{2}\}$, and ${DO}_{\Gamma\Gamma}=\min\{\sum_{i=1}^{N}\frac{\alpha_i^G}{2}, \sum_{i=1}^{N}\frac{\beta_i^G}{2},  \frac{\rho_U^2}{2}\}$, respectively.  The diversity order provides  deployment strategies for oceanic turbulence models and the beam-width of  optical transmissions. As such, beam-width can be adjusted sufficiently  to mitigate the effect of pointing errors.

\subsection{Average BER}
In this subsection, we derive the average BER for the proposed UWOC system. Considering IM/DD, the average BER can be obtained as \cite{dual_hop_turb2017}:
\begin{eqnarray} 
\overline{BER} = \frac{\delta}{2\Gamma(\phi)}\sum_{n=1}^{M'}q_n^{\phi}\int_{0}^{\infty} \gamma^{\phi-1} {\exp(-q_n \gamma)} F_{\gamma} (\gamma)   d\gamma
\label{eq:ber}
\end{eqnarray}
where the set $\{M', \delta, \phi, q_n\}$ can specify a variety of modulation schemes.

Using \eqref{eq:cdf_unified_hp} and substituting    $\exp(-q_n \gamma)=G_{0,1}^{1,0} \left[\begin{matrix} - \\ 0\end{matrix} \bigg| q_n \gamma\right]$ in \eqref{eq:ber} and reduce the Meijer's G-function into Fox-H function, we get  
\begin{eqnarray}
&\overline{BER}_{U}=\frac{\delta\rho_U^2}{2\Gamma(\phi)}\sum_{n=1}^{M'}q_n^{\phi}\prod_{i=1}^{N} \mathcal{A}_i\sum_{j=0}^{\mathcal{P}}\mathcal{B}_{i,j} \nonumber \\&\int_{0}^{\infty}\gamma^{\phi-1} H_{0,1}^{1,0} \left[\begin{matrix}  - \\ (0,1)\end{matrix} \bigg| q_n \gamma \right] H_{p+2,q+2}^{m+1,n+1}  \nonumber \\&\left[\begin{matrix} (1,1),(1+\rho_U^2) \\\{\mathcal{D}_{i_1,j}\}_{i_1=1}^N, (\rho_U^2,1), (0,1) \end{matrix} \bigg|\prod_{i_2=1}^{N} \frac{\mathcal{E}_{i_2,j}}{A_U} \sqrt{\frac{\gamma}{\bar{\gamma}_U}}\right]d\gamma \nonumber \\&
\label{eq:avg_BER_combined}
\end{eqnarray}

Finally, we apply the identity \cite[eq. $2.8.4$]{Kilbas} to get  the closed-form expression for the average BER of the multi-layer UWOC channel:
\begin{eqnarray}
&\overline{BER}_{U}=\frac{\delta\rho_U^2}{2\Gamma(\phi)}\sum_{n=1}^{M'}\prod_{i=1}^N\mathcal{A}_i\sum_{j=0}^{\mathcal{P}}\mathcal{B}_{i,j}H_{p+3,q+2}^{m+1,n+1}\nonumber \\& \left[\begin{matrix} (1,1),(1-\phi,\frac{1}{2}),(1+\rho_U^2) \\\{\mathcal{D}_{i_1,j}\}_{i_1=1}^N, (\rho_U^2,1), (0,1) \end{matrix} \bigg|\prod_{i_2=1}^{N} \frac{\mathcal{E}_{i_2,j}}{A_U\sqrt{q_n\bar{\gamma}_U}} \right]\nonumber \\&
\label{eq:avg_BER_combined_final}
\end{eqnarray}

Similar to the outage probability, the asymptotic expression for average BER at high SNR $\bar{\gamma}_U\to \infty$ can be derived as
\begin{eqnarray}
&\overline{BER}_{U}^{\infty}=\frac{\delta\rho_U^2}{2\Gamma(\phi)}\sum_{n=1}^{M'}\prod_{i=1}^N\mathcal{A}_i\sum_{j=0}^{\mathcal{P}}\mathcal{B}_{i,j}\sum_{k=1}^{m+1}\nonumber\\&\frac{\left(\frac{\mathcal{E}_{i_2,j}}{A_U\sqrt{q_n\bar{\gamma}_U}}\right)^{\frac{b_k}{\beta_k}}\prod_{j=1,j\neq k}^{m+1}\Gamma\left(b_j-b_k\frac{\beta_j}{\beta_k}\right)\Gamma\left(\frac{b_k}{\beta_k}\right)\Gamma\left(\phi+\frac{b_k}{2\beta_k}\right)}{\beta_k\Gamma\left(1+\rho_U^2-\frac{b_k}{\beta_k}\right)\Gamma\left(1+\frac{b_k}{\beta_k}\right)}
\label{eq:avg_BER_asymp}
\end{eqnarray}
where $b_j=b_k=[\mathcal{D}_{i_1,j}]_1$ and $\beta_j=\beta_k=[\mathcal{D}_{i_1,j}]_2$.

Thus, the dominant SNR terms of \eqref{eq:avg_BER_asymp} provides the diversity of proposed system as $\min\{\sum_{i=1}^{N}\frac{b_k}{2\beta_k}\}$, which is exactly same as obtained using  the outage probability.

\subsection{Ergodic Capacity}
The ergodic capacity for the underwater link $\overline{EC}_U$ is an important performance metric for  the design of communication systems and it can be defined as \cite{Nistazakis2009}:

\begin{eqnarray}
\overline{EC}_U&=\int\limits_{0}^{\infty} \log_2(1+\kappa\gamma) f_{\gamma_U}(\gamma) d\gamma
\label{eq:general_capacity_exp}
\end{eqnarray}
where $\kappa=\frac{e}{2 \pi}$ for IM/DD  and $\kappa=1$ for HD (heterodyne detection).

Using \eqref{eq:pdf_unified_hp} and substituting $ \log_2(1+\kappa\gamma)=1.44 G_{2,2}^{1,2} \left[\begin{matrix} 1,1 \\ 1,0\end{matrix} \bigg| \kappa\gamma \right]$ in \eqref{eq:general_capacity_exp} and reduce the Meijer's G-function into Fox-H function, we get

\begin{eqnarray}
&\overline{EC}_U=0.72\rho_U^2\prod_{i=1}^{N} \mathcal{A}_i\sum_{j=0}^{\mathcal{P}}\mathcal{B}_{i,j} \nonumber\\&\int_{0}^{\infty}\gamma^{-1} H_{2,2}^{1,2} \left[\begin{matrix} (1,1), (1,1) \\ (1,1), (0,1) \end{matrix} \bigg| \kappa\gamma \right]\nonumber \\&H_{p+1,q+1}^{m+1,n} \left[\begin{matrix} (1+\rho_U^2) \\\{\mathcal{D}_{i_1,j}\}_{i_1=1}^N, (\rho_U^2,1) \end{matrix} \bigg|\prod_{i_2=1}^{N} \frac{\mathcal{E}_{i_2,j}}{A_U} \sqrt{\frac{\gamma}{\bar{\gamma}_U}}\right]d\gamma \nonumber \\&
\label{eq:capacity_combine1}
\end{eqnarray}

Finally, we applying the identity \cite[eq. $2.8.4$]{Kilbas} to get  the closed-form expression for the ergodic capacity over the cascaded channel
\begin{eqnarray}
&\overline{EC}=0.72\rho_U^2\prod_{i=1}^{N} \mathcal{A}_i\sum_{j=0}^{\mathcal{P}}\mathcal{B}_{i,j}H_{p+3,q+3}^{m+3,n+1}\nonumber \\& \left[\begin{matrix} (0,\frac{1}{2}),(1,\frac{1}{2}),(1+\rho_U^2,1) \\\{\mathcal{D}_{i_1,j}\}_{i_1=1}^N,(\rho_U^2,1)(0,\frac{1}{2}),(0,\frac{1}{2}) \end{matrix} \bigg|\prod_{i_2=1}^{N} \frac{(\kappa\bar{\gamma}_U)^{-\frac{1}{2}}}{A_U}\right]\nonumber \\&
\label{eq:capacity_combine_final}
\end{eqnarray}
In what follows, we employ a terrestrial optical link to communicate with   the underwater transmission.
\section{Performance of Mixed TOWC-UWOC System}
In this section, we analyze the performance of a mixed TOWC-UWOC system when the fixed-gain AF relaying is applied.
We can use	\eqref{eq: System Model main eq} to express the end-to-end SNR  of the dual-hop system consisting of  TOWC and UWOC links \cite{Hasna2004}:
\begin{equation}
	\label{eq:af}
	{\gamma} = \frac{\gamma_{T}\gamma_{U}}{\gamma_{U}+C}
\end{equation}
where $C$ a constant for the fixed-gain AF relaying protocol. Standard transformation of random variables in \eqref{eq:af} leads to th the PDF of SNR for the fixed-gain  AF relayed system as
\begin{equation} \label{eq:snr_pdf_eqn_af_gen}
	f_{\gamma^{}}(\gamma) = \int_{0}^{\infty} {f_{\gamma_{T}}\left(\frac{\gamma(x + C)}{x}\right)} {f_{\gamma_{U}}(x)} \frac{x + C}{{x}} {dx}
\end{equation}  
where $f_{\gamma_{T}}(\gamma)$ and $f_{\gamma_{U}}(\gamma)$ are the PDF of SNR for TOWC link and UWOC link, respectively. 

We use \eqref{eq:pdf_fpt_twoc} and \eqref{eq:pdf_unified_hp} in 	\eqref{eq:snr_pdf_eqn_af_gen} to develop the PDF of the SNR for the mixed link:
\begin{my_lemma}\label{lemma:pdfcdf_af}
	The PDF and CDF of the end-to-end SNR for the fixed-gain AF relay-assisted mixed TWOC-UWOC system are given as
	\begin{flalign} \label{eq:snr_pdf_eqn_af}
		&f_{\gamma}(\gamma) = \frac{z^k\rho_T^2A^{\rm{mg}}}{4 \gamma}  \sum_{m_1=1}^{\beta^M}b^M_{m_1}  \prod_{i=1}^{N}\frac{\rho_U^2\mathcal{A}_i}{2} \sum_{j=0}^{\mathcal{P}}\mathcal{B}_{i,j}  \nonumber \\ &H_{1,0:3+k,k+2;1,m+2}^{0,1:0,3+k;m+2,0} \Bigg[\begin{array}{c} \frac{(g^M\beta^M+\Omega^M)A_T\sqrt{\bar{\gamma}_T}}{\alpha^M\beta^M \sqrt{\gamma}} \\ \frac{\mathcal{E}_{i_2,j}}{A_U} \sqrt{\frac{1}{\bar{\gamma}_U}}  \end{array} \bigg\vert \begin{array}{c} U_1\\ V_1\end{array} \Bigg]
	\end{flalign}
	where $U_1=\{(1:\frac{1}{2},\frac{1}{2}):(1-\rho_T^2,1);(1-\alpha^M,1);(1-m_1,1);\{(1-z,1)\}^k_1;(1+\rho_U^2,1)\}$ and $V_1=\{- : (-\rho_T^2,1);(-z,1);(1,\frac{1}{2});\{\mathcal{D}_{i_1,j}\}_{i_1=1}^{N}; (\rho_U^{2},1);(0,\frac{1}{2})\}$.
	
		\begin{flalign} \label{eq:snr_cdf_eqn_af}
		&F_{\gamma}(\gamma) = \frac{z^k\rho_T^2A^{\rm{mg}}}{4}  \sum_{m_1=1}^{\beta^M}b^M_{m_1}  \prod_{i=1}^{N}\frac{\rho_U^2\mathcal{A}_i}{2} \sum_{j=0}^{\mathcal{P}}\mathcal{B}_{i,j}  \nonumber \\ &H_{1,0:4+k,3+k;1,2+m}^{0,1:0,4+k;2+m,0} \Bigg[\begin{array}{c} \frac{(g^M\beta^M+\Omega^M)A_T\sqrt{\bar{\gamma}_T}}{\alpha^M\beta^M \sqrt{\gamma}} \\ \frac{\mathcal{E}_{i_2,j}}{A_U} \sqrt{\frac{1}{\bar{\gamma}_U}}  \end{array} \big\vert \begin{array}{c} U_2\\ V_2\end{array} \Bigg]
		\end{flalign}
	where $U_2=\{(1:\frac{1}{2},\frac{1}{2}):(1-\rho_T^2,1);(1-\alpha^M,1);(1-m_1,1);\{(1-z,1)\}^k_1;(1,\frac{1}{2});(1+\rho_U^2,1)\}$ and $V_2=\{- :(-\rho_T^2,1);(-z,1);(1,\frac{1}{2});(0,\frac{1}{2}); \{\mathcal{D}_{i_1,j}\}_{i_1=1}^{N}; (\rho_U^{2},1);(0,\frac{1}{2})\}$
		
\end{my_lemma}
\begin{table*}[tp]	
	\renewcommand{\arraystretch}{01}
	\caption{Simulation Parameters}
	\label{table:simulation_parameters}
	\centering
	\begin{tabular}{|p{7cm}|c|c|c|}
		\hline 	
		Transmitted optical power &$P_t$ & $-10$ to $60$ \mbox{dBm} \\ \hline
		
		AWGN variance &$\sigma_w^2$ & $10^{-14}~\rm {A^2/GHz}$ \\ \hline	
		Total link distance &$l$ & $l=(l_T+l_U)=(400+50)$ \mbox{m}\\ \hline
		Extinction coefficient& $\alpha$ & $0.056$ \\ \hline
		Shape parameter of foggy channel& $k$ & \{13.12, 12.06\} \\ \hline
		Scale parameter of foggy channel& $\beta^f$ & \{2, 5\} \\ \hline
		Pointing errors parameters& $A$, $\rho^2$ & $0.0032$, $\{1, 6\}$
		\\ \hline
		Mal\'aga distribution parameters \cite{Chapala2021}
		& $\alpha^M$ &  $\{4.5916, 2.3378, 1.4321\}$\\& $\beta^M$ &  $\{7.0941, 4.5323, 3.4948\}$ 	\\ \hline
		& $\{a_i\}_{i=1}^5$ &  $\{0.6302, 1.0750, 1.0173, 0.7598, 1.0990\}$\\GG distribution parameters \cite{Vahid2018}& $\{d_i\}_{i=1}^5$ &  $\{1.1780, 3.2048, 1.6668, 2.3270, 4.5550\}$ \\ & $\{p_i\}_{i=1}^5$ &  $\{0.8444, 2.9222, 1.0380, 1.4353, 4.6208\}$\\ \hline EGG distribution parameters \cite{Zedini2019}
		& $\{\omega_i\}_{i=1}^2$ &  $\{0.1770, 0.4589\}$\\& $\{\lambda_i\}_{i=1}^2$ &  $\{0.4687, 0.3449\}$ 	\\ \hline
		& $\{\omega_i\}_{i=1}^5$ &  $\{0.2130, 0.2108, 0.1807, 0.1665, 0.4589\}$\\& $\{\lambda_i\}_{i=1}^5$ &  $\{0.3291, 0.2694, 0.1641, 0.1207, 0.3449\}$  \\ EGG distribution parameters (For Fig.~\ref{avg_ber_out_prob_fpt}(a) and Fig.~\ref{avg_ber_out_prob_fpt}(b)) \cite{Zedini2019}& $\{\frac{d_i}{p_i}\}_{i=1}^5$ &  $\{1.4299, 0.6020, 0.2334, 0.1559, 1.0421\}$ \\ & $\{a_i\}_{i=1}^5$ &  $\{1.1817, 1.2795, 1.4201, 1.5216, 1.5768\}$\\ & $\{p_i\}_{i=1}^5$ &  $\{17.1984, 21.1611, 22.5924, 22.8754, 35.9424\}$ \\ \hline
		EW distribution parameters \cite{Vahid2018}& $\alpha^{E}$, $\beta^{E}$, $\eta^{E}$ & $2.50$, $0.70$, $0.50$\\ \hline
		$\Gamma\Gamma$ distribution parameters \cite{Vahid2018}& $\alpha^{G}$, $\beta^{G}$ & $5$, $1.18$\\ \hline
		Modulation parameters& $M'$, $\delta$, $\phi$, $q_n$ & $1$, $1$, $\frac{1}{2}$, $\frac{1}{2}$ 
		
		\\ \hline	
	\end{tabular}
	\label{Simulation_Parameters}	 
\end{table*}

\begin{figure*}[t]
	\centering
	\subfigure[Single-layer ($N=1$) with $\rho_U^2=1$.]
	{\includegraphics[scale=0.30]{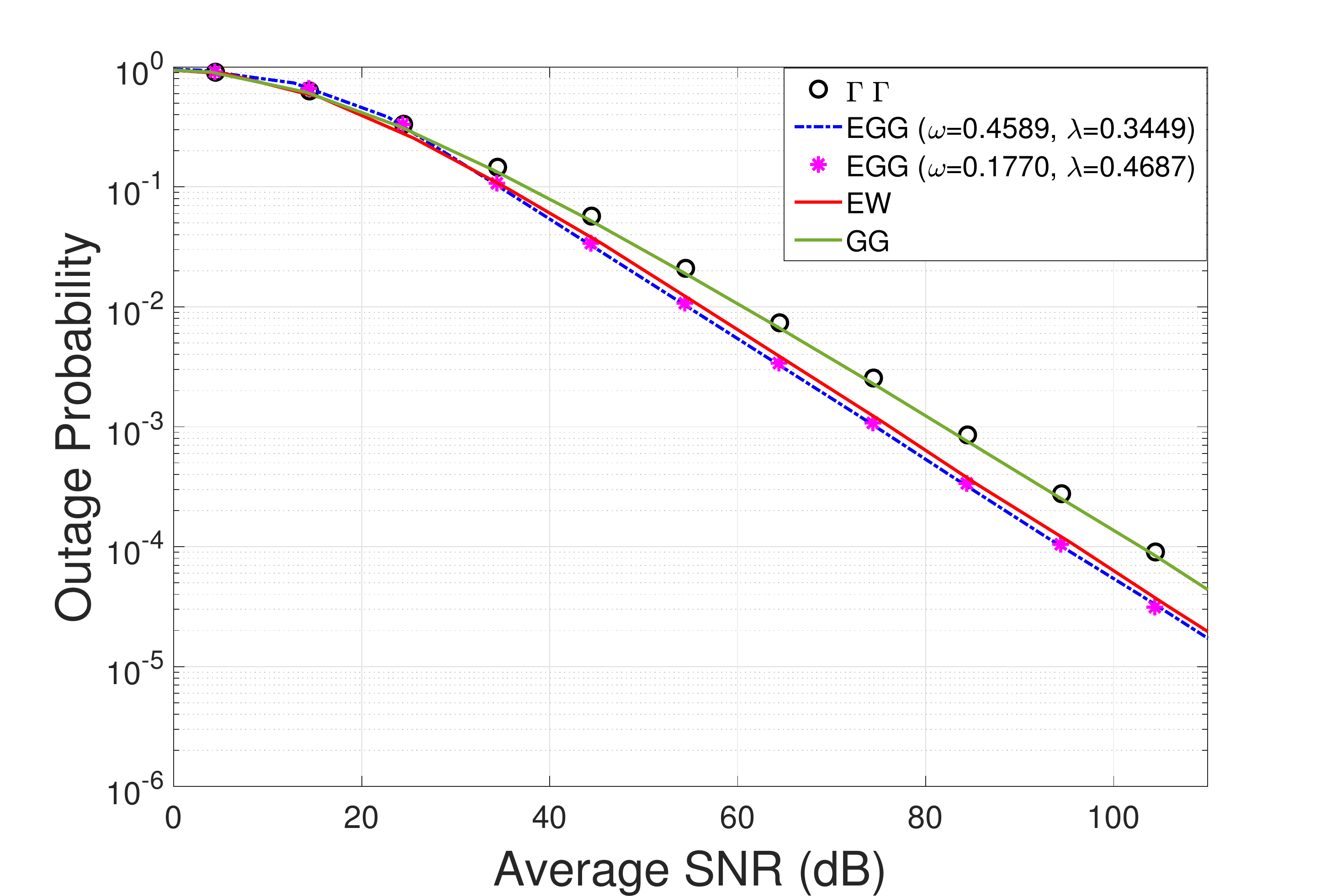}}
	\subfigure[Multi-layer ($N=5$) with GG oceaninc turbulence.]
	{\includegraphics[scale=0.30]{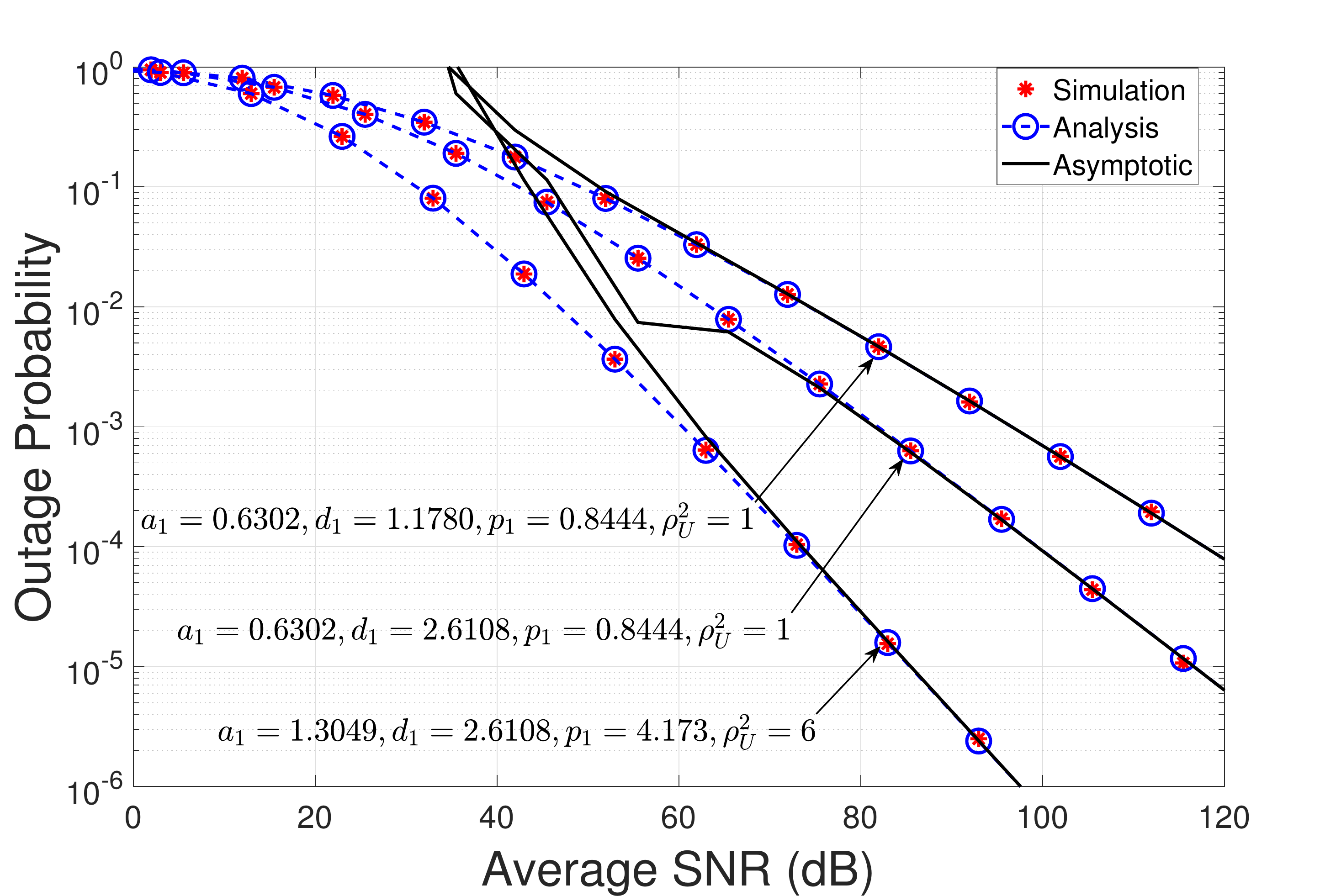}}
	\caption{Outage probability performance.}
	\label{out_prob_n1_5}
\end{figure*}
 
 \begin{figure*}[t]
 	\centering
 	\subfigure[Single-layer ($N=1$) for various oceaninc turbulence.]
 	{\includegraphics[scale=0.30]{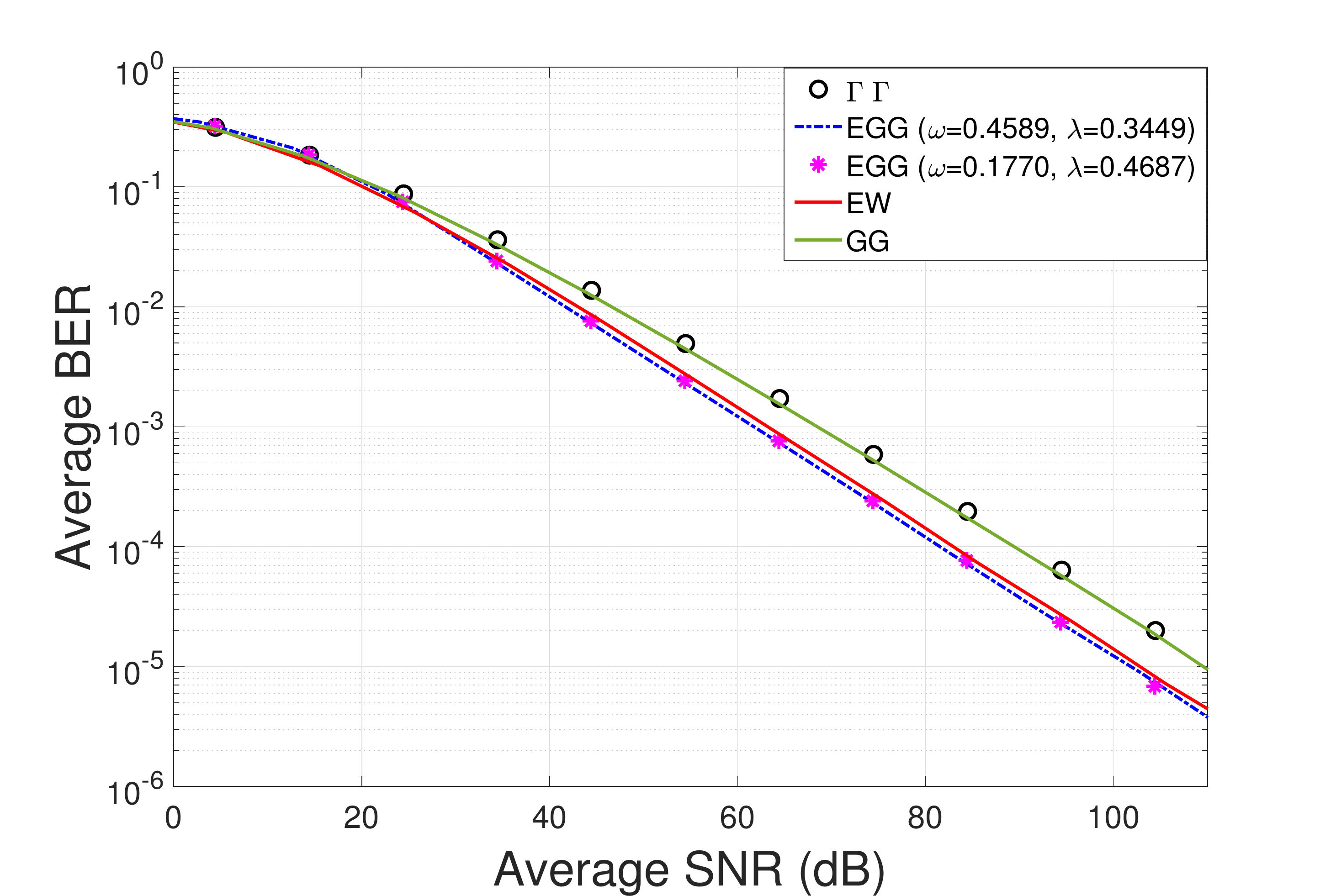}}
 	\subfigure[Multi-layer ($N=5$) with GG oceaninc turbulence and $\rho_U^2=6$.]
 	{\includegraphics[scale=0.30]{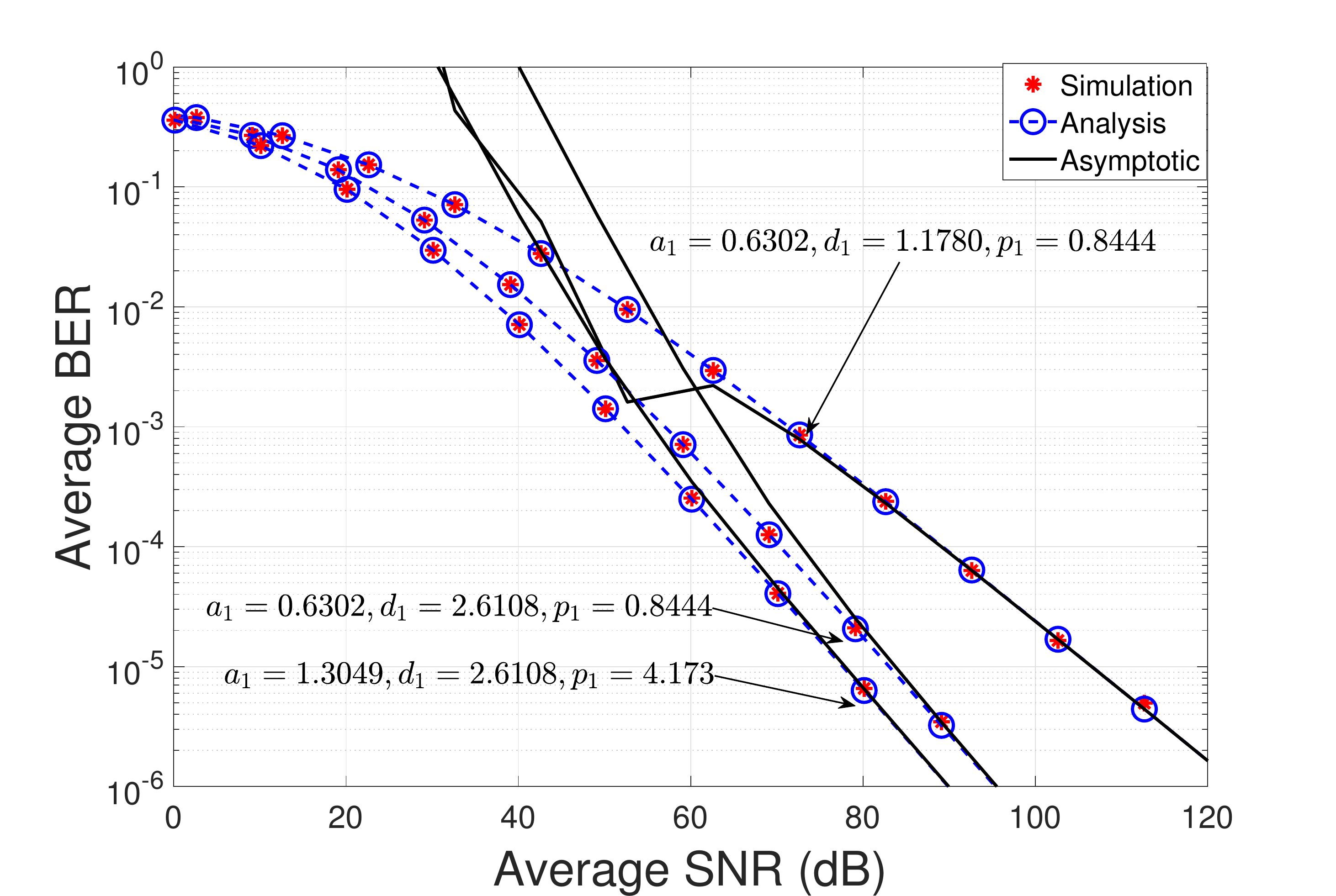}}
 	\caption{Average BER performance.}
 	\label{avg_ber_n1_5}
 \end{figure*}
 
 \begin{figure*}[t]
 	\centering
 	\subfigure[Single-layer ($N=1$) for various oceaninc turbulence.]
 	{\includegraphics[scale=0.30]{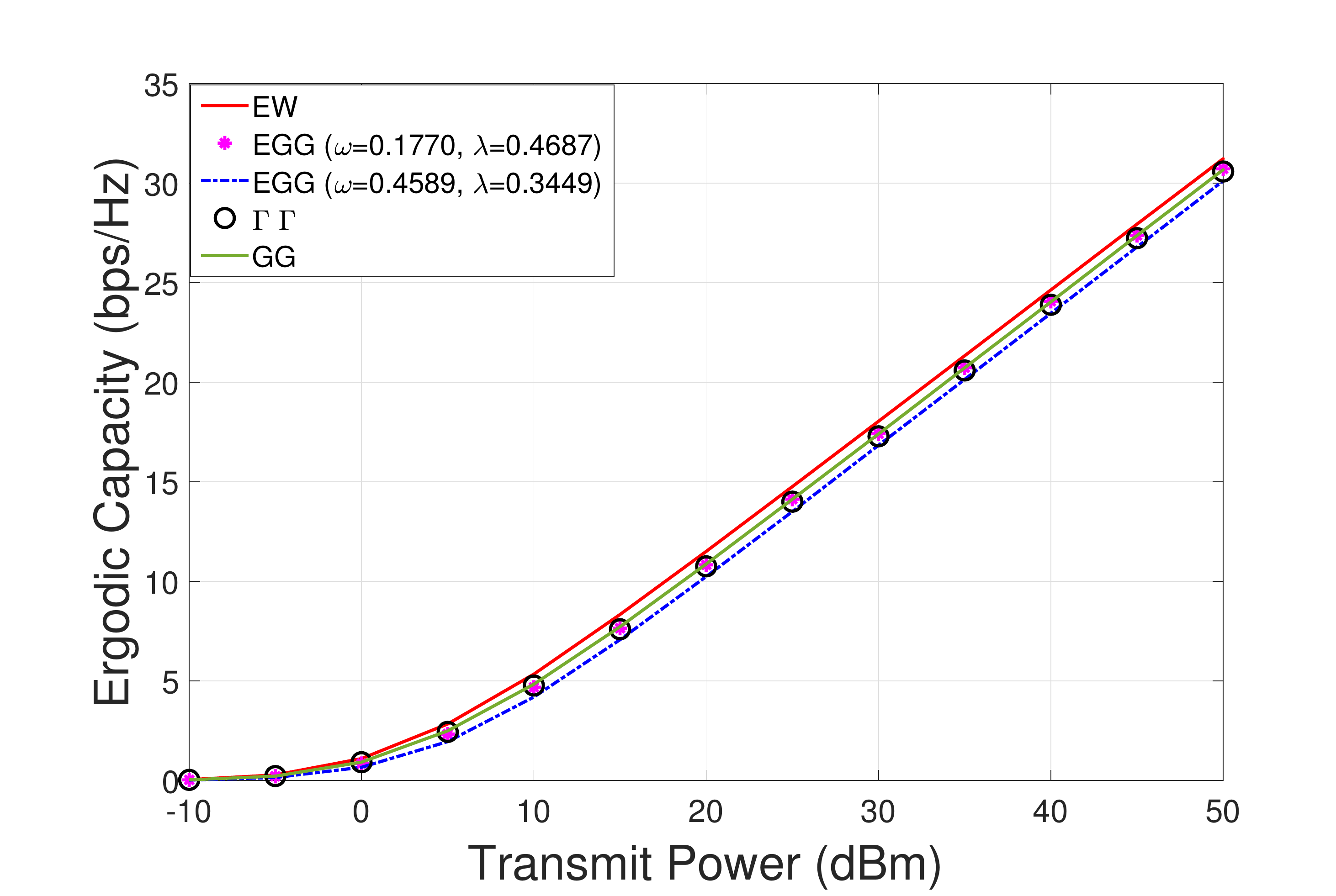}}
 	\subfigure[Multi-layer ($N=5$) with GG oceaninc turbulence.]
 	{\includegraphics[scale=0.30]{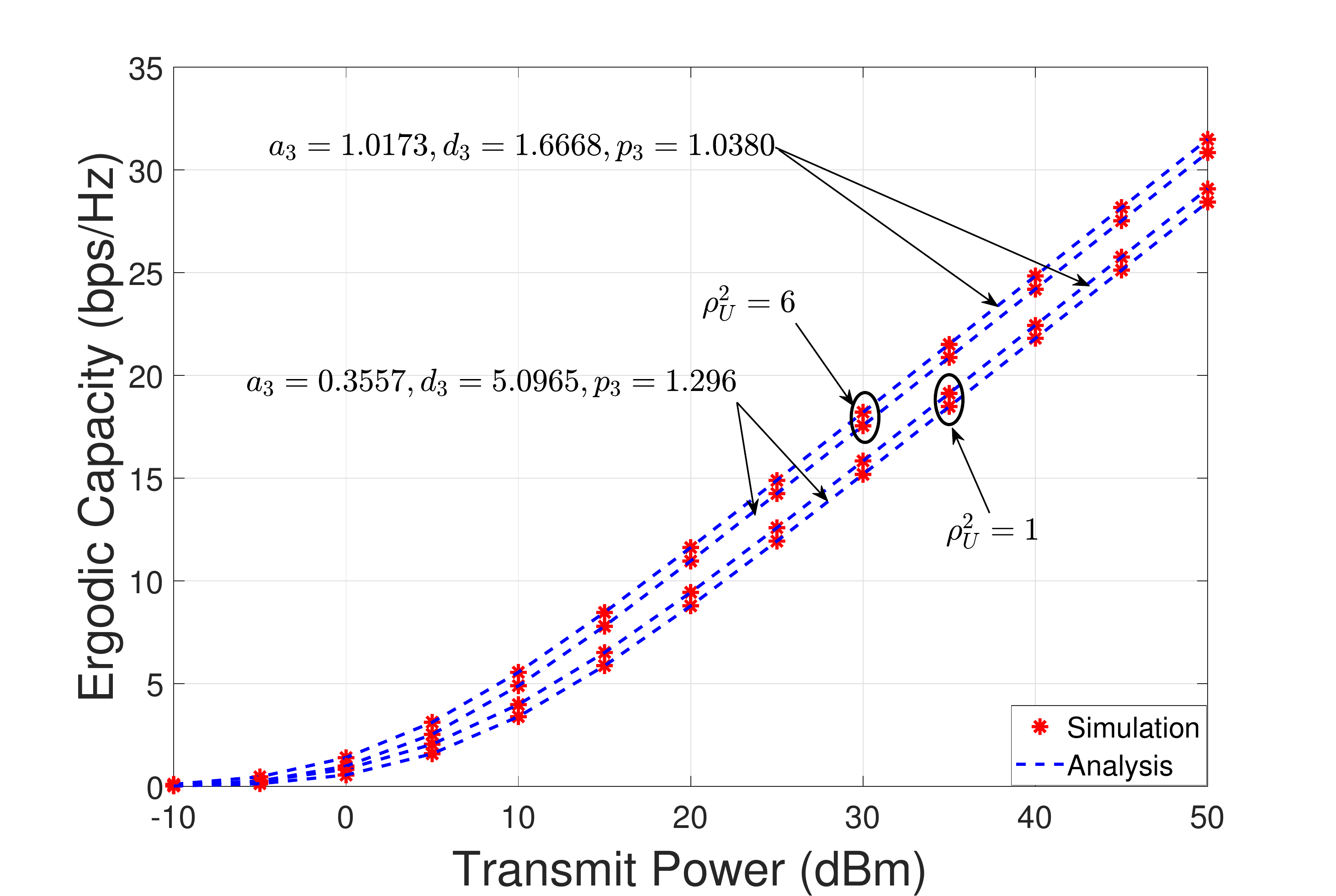}}
 	\caption{Ergodic capacity performance.}
 	\label{erg_cap_n1_5}
 \end{figure*}
 
 \begin{figure*}[t]
 	\centering
 	\subfigure[Outage probability.]
 	{\includegraphics[scale=0.30]{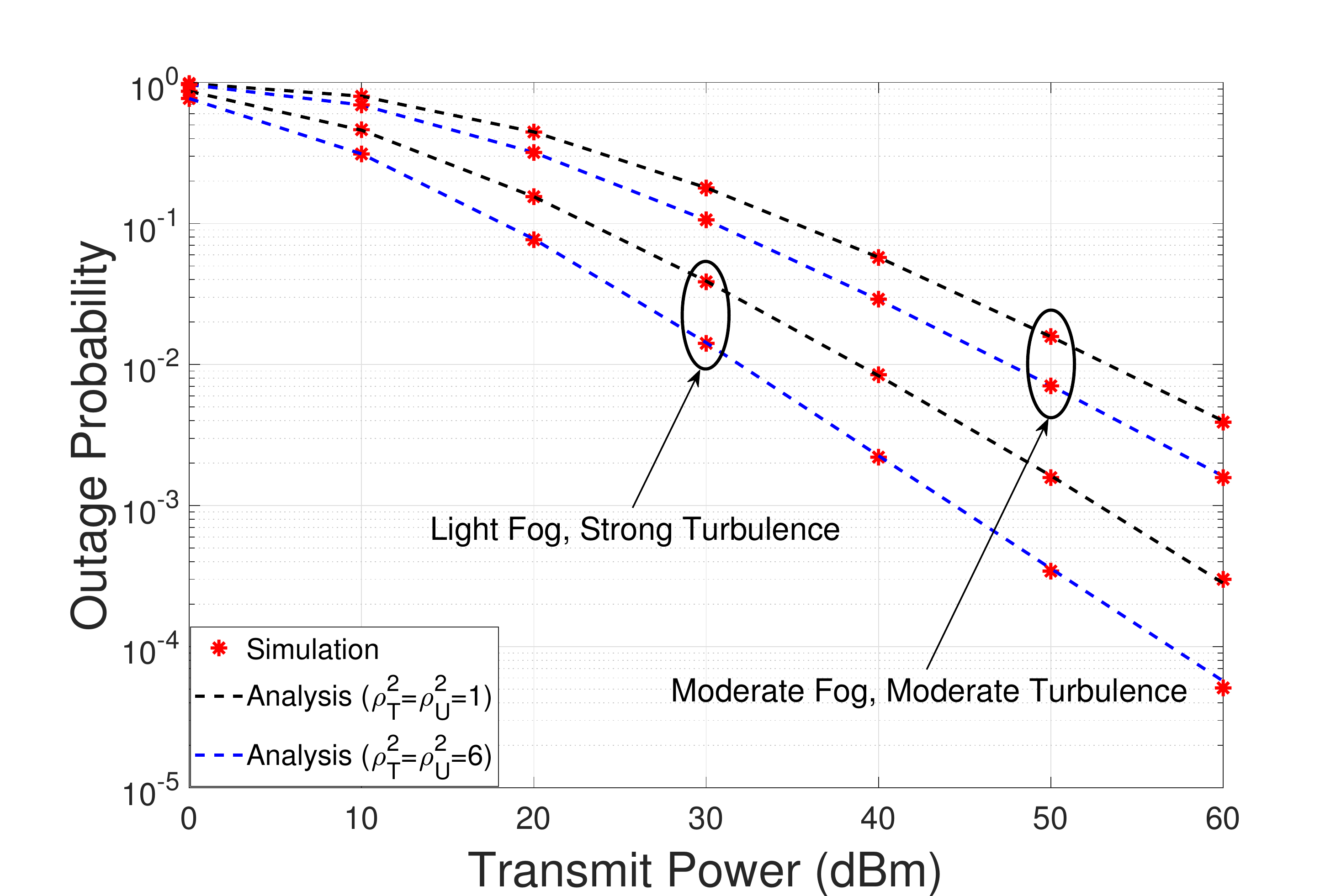}}
 	\subfigure[Average BER with $\rho_T^2=6$, and $\rho_U^2=6$.]
 	{\includegraphics[scale=0.30]{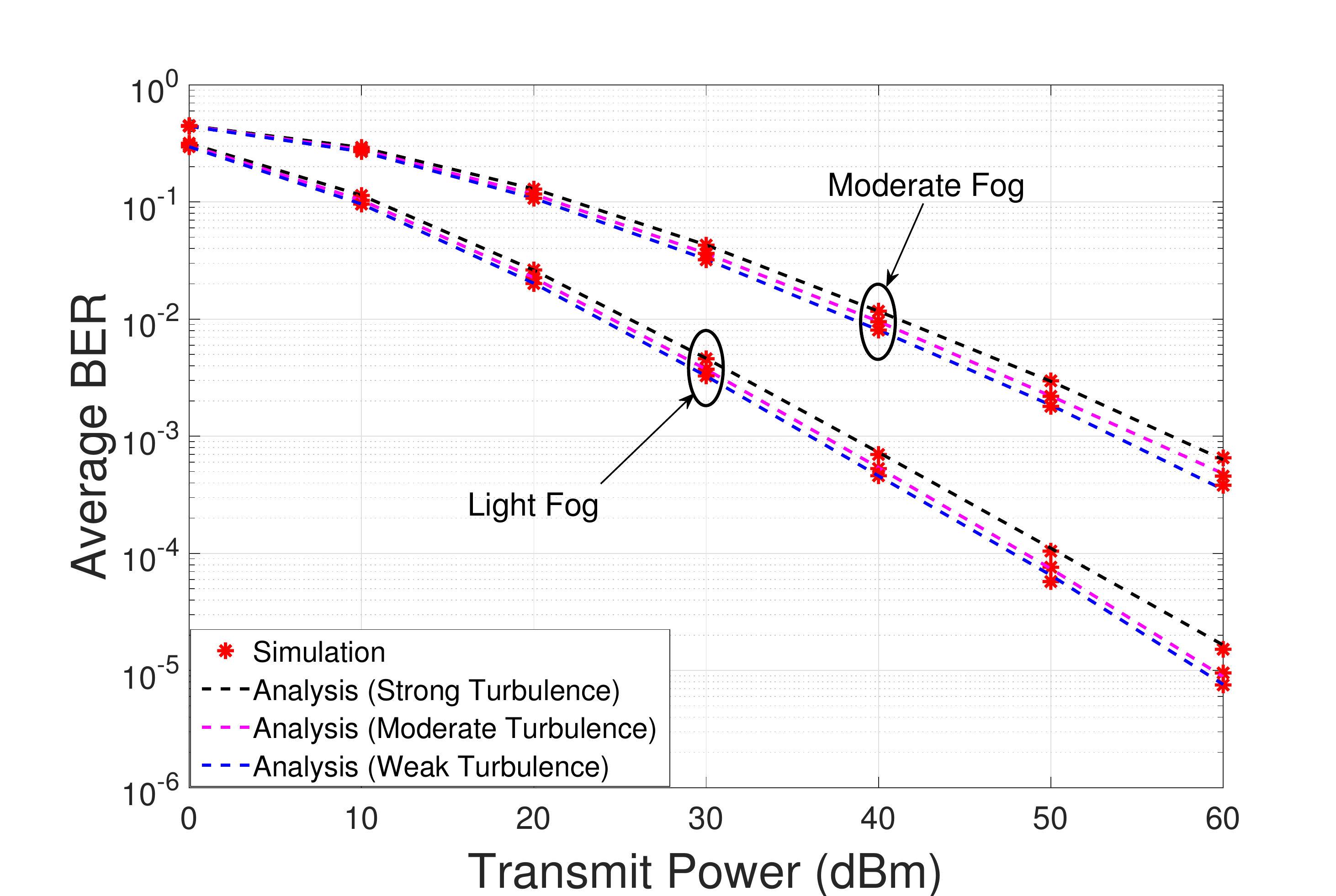}}
 	\caption{Outage and error performance of mixed TWOC and multi-layer UWOC with EGG oceanic turbulence.}
 	\label{avg_ber_out_prob_fpt}
 \end{figure*}
\begin{IEEEproof} 
	See Appendix C.
\end{IEEEproof}
There are standard routines to compute bivariate Fox-H function in MATLAB. Further, we can use the CDF in \eqref{eq:snr_cdf_eqn_af} to derive the outage probability of the mixed TWOC-UWOC system as $P_{\text{out}}=F_{\gamma}(\gamma_{\text{th}})$ given a threshold SNR $\gamma_{\rm th}$.

Finally, we use \eqref{eq:snr_cdf_eqn_af} in \eqref{eq:ber} to find the average BER of the mixed TWOC-UWOC system as
\begin{align} \label{eq:avg_ber_eqn_af_1}
	&\overline{BER} =  \frac{\delta}{2\Gamma(\phi)}\sum_{n=1}^{M'}q_n^{\phi} \frac{z^k\rho_T^2A^{\rm{mg}}}{4}  \sum_{m_1=1}^{\beta^M}b^M_{m_1}  \bigg(\frac{1}{2\pi \J}\bigg)^{2} \nonumber\\&\int_{\mathcal{L}_{1}}\int_{\mathcal{L}_{2}} \prod_{i=1}^{N}\frac{\rho_U^2\mathcal{A}_i}{2} \sum_{j=0}^{\mathcal{P}}\mathcal{B}_{i,j} \bigg(\frac{(g^M\beta^M+\Omega^M)A_T\sqrt{\bar{\gamma}_T}}{\alpha^M\beta^M \sqrt{\gamma}}\bigg)^{-n_{1}}\nonumber \\ & \frac{\Gamma(\rho_T^{2}-n_{1})\Gamma(\alpha^M-n_{1})\Gamma(m_1-n_{1})[\Gamma(z-n_{1})]^k_1}{\Gamma(\rho_T^{2}+1-n_{1})[\Gamma(z+1-n_{1})]^k_1\Gamma(-\frac{n_{1}}{2})} \nonumber \\ &\bigg(\frac{\mathcal{E}_{i_2,j}}{A_U} \sqrt{\frac{1}{\bar{\gamma}_U}}\bigg)^{-n_{2}}  \frac{\prod_{j=1}^{m} [\Gamma(\mathcal{D}_{i,j})]^{N}_{i=1}\Gamma(\rho_{U}^{2}+n_{2})\Gamma(\frac{n_{2}}{2})}{\Gamma(\rho_{U}^{2}+1+n_{2})}   \nonumber\\&\Gamma\left(-\frac{n_{1}}{2}-\frac{n_{2}}{2}\right) \bigg(\int_{0}^{\infty} \gamma^{\phi-\frac{n_{1}}{2}-1}\exp(-q_n \gamma)\diff \gamma\bigg) \diff n_{1} \diff n_{2} 
\end{align}
Solving  the inner integral as $\int_{0}^{\infty} \gamma^{\phi-\frac{ n_{1}}{2}-1}\exp(-q_n \gamma)\diff \gamma=\Gamma\left(\phi-\frac{n_1}{2}\right)q^{-\phi+\frac{n_1}{2}}_n$ and applying   \cite{Mittal_1972}, we get average BER of the mixed TWOC-UWOC system involving the bivariate Fox-H function:
\begin{align} \label{eq:avg_ber_eqn_af}
&\overline{BER}=  \frac{\delta}{2\Gamma(\phi)}\sum_{n=1}^{M'}\frac{z^k\rho_T^2A^{\rm{mg}}}{4}  \sum_{m_1=1}^{\beta^M}b^M_{m_1}  \prod_{i=1}^{N}\frac{\rho_U^2\mathcal{A}_i}{2} \sum_{j=0}^{\mathcal{P}}\mathcal{B}_{i,j}  \nonumber \\ & H_{1,0:5+k,3+k;1,2+m}^{0,1:0,5+k;2+m,0} \Bigg[\begin{array}{c} \frac{(g^M\beta^M+\Omega^M)A_T\sqrt{\bar{\gamma}_T}}{\alpha^M\beta^M \sqrt{q_n}} \\ \frac{\mathcal{E}_{i_2,j}}{A_U} \sqrt{\frac{1}{\bar{\gamma}_U}}  \end{array} \big\vert \begin{array}{c} U_3\\ V_3\end{array} \Bigg] 
\end{align}
where $U_3=\{(1:\frac{1}{2},\frac{1}{2}):(1-\rho_T^2,1);(1-\alpha^M,1);(1-m_1,1);\{(1-z,1)\}^k_1;(1,\frac{1}{2});\left(1-\phi,\frac{1}{2}\right);(1+\rho_U^2,1)\}$ and $V_3=\{- :(-\rho_T^2,1);(-z,1);(1,\frac{1}{2});(0,\frac{1}{2}); \{\mathcal{D}_{i_1,j}\}_{i_1=1}^{N}; (\rho_U^{2},1);(0,\frac{1}{2})\}$
It should be mentioned  that bivariate Fox-H function has been extensively  analyzing the fixed-gain AF relaying over complicated fading models \cite{Yang2021}.

\section{Simulation and numerical analysis}
In this section, we demonstrate the performance of multi-layer vertical UWOC system over-various oceanic turbulence conditions and the mixed TOWC-UOWC transmission. We also compare the performance of the multi-layer UWOC with the single-layer ($N=1$) approximation. We use Monte-Carlo (MC) simulation (averaged over $10^7$ channel realizations) to validate the derived analytical expressions. Further,  the asymptotic expression of outage probability and average BER converges with analysis and simulation results in the high SNR regime. We use standard inbuilt MATLAB and Mathematica libraries to calculate Meijer's G and Fox’s H-function, respectively. Since there is no measurement data to confirm the variation of distribution parameters with distance, we illustrate the performance by considering vertical underwater link length $l_U=50$ {\rm{m}} with $N=5$ layers, and the thickness of each layer is assumed to be $10$ {\rm{m}}.   We use standard simulation parameters and measurement-based parametric data for EGG, GG, EW, and $\Gamma \Gamma$, as given in Table \ref{Simulation_Parameters}.

First, we demonstrate the outage probability performance of the considered UWOC system in Fig.~\ref{out_prob_n1_5}. It can be seen from Fig.~\ref{out_prob_n1_5}(a) that the outage probability for a specific oceanic turbulence condition using different statistical models is similar for the  single-layer ($N=1$) case validating the equivalence of different models using PDF plots, as demonstrated comprehensively in \cite{Vahid2018}. We used two different   EGG turbulence  since  $\lambda$ and $\omega$ parameters is not available for the same experimental scenario. The figure shows that an acceptable operating outage probability of $10^{-3}$ can be achieved with an average SNR of $80$\mbox{dB}.  To demonstrate the multi-layer performance, we consider the GG model (as depicted in Fig.~\ref{out_prob_n1_5}(b) ), which  excellently fits the experimental data for a wide range of oceanic turbulence, as observed in \cite{Vahid2018}. Comparing Fig.~\ref{out_prob_n1_5}(a) and Fig.~\ref{out_prob_n1_5}(b) with $\rho_U^2=1$ plots, it can be seen that the single-layer model underestimates the oceanic turbulence concerning the $N=5$ layers case justifying the use of multi-layer modeling for UWOC transmissions. Further,  Fig.~\ref{out_prob_n1_5}(b) shows that the outage performance of the system improves with an increase in the values of GG distribution parameters ($a$, $d$, and $p$) and a decrease in  pointing errors (i.e., higher $\rho_U^2$). In the first plot of Fig.~\ref{out_prob_n1_5}(b), we consider the pointing errors parameter ($\rho_U^2=1$) and the GG distribution parameters ($a$, $d$, and $p$) as given in Table \ref{Simulation_Parameters}. The diversity order ${DO}_{\rm out}=\min\{\sum_{i=1}^{N}\frac{d_i}{2}, \frac{\rho_U^2}{2}\}$ for the top and middle plots in  Fig.~\ref{out_prob_n1_5}(b) are given by $\min\{6.4658, 0.5\}$ and $\min\{7.1822, 0.5\}$, respectively. It can be clearly observed that the diversity order is dependent on the pointing error parameter ($\rho_U^2$) since the slope does not change  with the oceanic channel parameter $d_i$. Further, in the third plot,  the diversity order becomes $\min\{7.1822, 3\}$, demonstrating a  change of slope with  $\rho_U^2$, thus confirming our diversity order analysis. 

Next, we present the average BER performance of the single-layer system in Fig.~\ref{avg_ber_n1_5}(a) and multi-layer UWOC in Fig.~\ref{avg_ber_n1_5}(b). Similar to the outage probability, the average BER of the system provides similar observations with respect to the comparison of single-layer and multi-layer models and  follows a similar trend  with turbulent channel and pointing error parameters, as shown in Fig.~\ref{avg_ber_n1_5}(b). It is evident from the plots that the average BER of the system improves by almost ten times if we increase the channel parameter $d_i$ from $1.1780$ to $2.6108$ at average SNR of $80$ \mbox{dB}. The diversity order follows a similar analysis as that of the outage probability, which can be confirmed by observing  the  slope change among the plots.

In Fig.~\ref{erg_cap_n1_5}(a) and Fig.~\ref{erg_cap_n1_5}(b), we plot ergodic capacity performance for the single-layer ($N=1$) and multi-layer ($N=5$) UWOC systems, respectively. The ergodic capacity for different UWOC models is almost equal for the given oceanic conditions, as shown in Fig.~\ref{erg_cap_n1_5}(a). Further, the ergodic capacity is around $10$\mbox{bits/sec/Hz} at a nominal transmit power of $20$\mbox{dBm}.   In Fig.~\ref{erg_cap_n1_5}(b), the effect of GG parameters is demonstrated for the multi-layer UWOC transmission. The figure shows that the ergodic
capacity increases by almost $3 {\rm{bits/sec/Hz}}$ if the pointing error parameter $\rho_U^2$ increases from $1$ to $6$ for the given turbulent  parameters. It can also been seen that the ergodic capacity increases by almost $1 {\rm{bits/sec/Hz}}$ if we change the oceanic turbulence parameters ($a_3=0.3557$, $d_3=5.0965$, and $p_3=1.296$) for the both  $\rho_U^2=1$ and $\rho_U^2=6$.
 
Finally, we demonstrate the outage probability and average BER performance of the mixed TWOC and multi-layer ($N=5$) UWOC system in Fig.~\ref{avg_ber_out_prob_fpt}(a) and Fig.~\ref{avg_ber_out_prob_fpt}(b), respectively. We consider terrestrial link distance $l_T= 400$ \rm{m},   underwater link length $l_U=50$ \rm{m} with $N=5$ layers each with  EGG  oceanic turbulence. In Fig.~\ref{avg_ber_out_prob_fpt}(a), we plot the outage probability considering light fog with weak turbulence and moderate fog with moderate turbulence for two different pointing errors parameters $\rho_T^2=\rho_U^2=\{1,6\}$. The effect of fog density, the intensity of atmospheric turbulence, and pointing errors are clearly visible. It can be seen from Fig.~\ref{avg_ber_out_prob_fpt}(a), that $20$ \mbox{dBm} more transmit power is required to achieve the same outage probability with  moderate fog   as compared with light fog. Further, the penalty for strong pointing errors is $10$\mbox{dBm} for the same foggy and atmospheric  conditions. In Fig.~\ref{avg_ber_out_prob_fpt}(b), we demonstrate the effect of different (weak, moderate, and strong) atmospheric turbulence on the  average BER performance for both light and moderate foggy conditions. The figure shows that the effect of atmospheric turbulence  on the average BER performance is less as compared with performance degradation  due to the fog.

\section{Conclusions and Future Work} 
We presented unified  performance analysis of the UWOC system considering the vertical underwater link as a multi-layer cascaded channel considering i.ni.d. GG, EGG, EW, and $\Gamma\Gamma$ oceanic turbulence channels. We analyzed the system performance by deriving analytical expressions of the PDF and CDF of the end-to-end  SNR, and developed  outage probability, average BER, and ergodic capacity  under the combined effect of cascaded oceanic turbulence and pointing errors in terms of Meijer-G and Fox-H functions.  We  provided the asymptotic expressions using Gamma functions for the outage probability and average BER to determine the diversity order of the considered system. We also  employed the fixed-gain AF relaying to integrate the terrestrial OWC transmission subjected to the combined effect of  generalized Mal\'aga atmospheric turbulence, fog-induced random path gain, and pointing errors to communicate with the UWOC link. We analyzed the performance of the mixed link using outage probability and average BER involving bivariate Fox H-function. Simulation results showed a  performance gap when the  single-layer approximation was  compared with the multi-layer model. The proposed analysis would be  helpful for an efficient deployment for UWOC under various oceanic conditions.
The existing measurement data and statistical model  do not consider depth dependency for oceanic turbulence. It would be interesting to investigate the applicability of the proposed analysis using the channel measurement data considering  ocean stratification for the UWOC system.

\section*{Acknowledgment}
We wish to thank Mr. Suhrid Das for his work on the  multi-layer GG oceanic turbulence model.

\section*{Appendix A: Proof of Theorem 1}
 First, we find the PDF of $h_{\rm GG}=\prod_{i=1}^{N}h_{i}$, where $h_i$, $i=1,2, \cdots, N$ denote  i.ni.d GG random variables. Substituting \eqref{eq:h_gg} in \eqref{eq:nth_order_moment_of_hi} and applying the identity $\int_{0}^{\infty} t^a \exp(-bt^c)dt=\frac{\Gamma\left(\frac{1+a}{c}\right)}{c b^{\frac{1+a}{c}}}$ \cite[pp. $347$, eq. $3.381.10$]{Zwillinger2014}, we get the $n$-th order moment of $h_{c}$ as:
\begin{equation}
	\mathbb{E}[h_{\rm GG}^n]=\prod_{i=1}^{N} \frac{\Gamma\left(\frac{n+d_i}{p_i}\right)}{a_i^{-n}\Gamma\left(\frac{d_i}{p_i}\right)}
	\label{eq:nth_order_moment_of_h_gen_g}
\end{equation}
We use \eqref{eq:nth_order_moment_of_h_gen_g} in \eqref{eq:Nth_order_with_Inverse_Mellin} and apply the definition of Fox H-function to get the PDF of $h_c$ for the generalized Gamma turbulent channel as 
\begin{eqnarray}
	&f_{h_{GG}}(x)=\prod_{i=1}^{N} \frac{1}{\Gamma\left(\frac{d_i}{p_i}\right)}\nonumber \\& \frac{1}{x}H_{0,N}^{N,0} \left[\begin{matrix} - \\\left\{\left(\frac{d_{i_1}}{p_{i_1}},\frac{1}{p_{i_1}} \right)\right\}_{i_1=1}^{N}  \end{matrix} \bigg|\prod_{i_2=1}^{N} \left(\frac{x}{a_{i_2}}\right)\right]
	\label{eq:pdf_N_gen_gamma}
\end{eqnarray}  
Next, we find the PDF of $h_{\rm EGG}=\prod_{i=1}^{N}h_{i}$, where $h_i$, $i=1,2, \cdots, N$ denote  i.ni.d EGG random variables. Substituting \eqref{pdf_uw} in \eqref{eq:nth_order_moment_of_hi}, we get the $n$-th order moment of $h_{\rm EGG}$ as:
\begin{eqnarray}
	\mathbb{E}[h_{\rm EGG}^n]=\prod_{i=1}^{N} \Bigg(\omega_i \lambda_i^n\Gamma(1+n) +\frac{(1-\omega_i)\Gamma\left(\frac{n+d_i}{p_i}\right)}{a_i^{-n}\Gamma\left(\frac{d_i}{p_i}\right)}\Bigg)
	\label{eq:nth_order_moment_of_h_egg_1}
\end{eqnarray}	
Using \eqref{eq:nth_order_moment_of_h_egg_1} in \eqref{eq:Nth_order_with_Inverse_Mellin} and applying the definition of Fox H-function to get the PDF of $h_{\rm EGG}$  for the EGG oceanic turbulent channel:
\begin{eqnarray}
	&f_{h_{\rm EGG}}(x)=\prod_{i=1}^{N} \Bigg(\frac{\omega_i}{x}H_{0,N}^{N,0} \left[\begin{matrix} - \\ \{(1,1)\}_{i_1=1}^{N} \end{matrix} \bigg| \prod_{i_2=1}^{N}\frac{x}{\lambda_{i_2}}\right]+\nonumber\\&\frac{(1-\omega_i) }{ \Gamma\left(\frac{d_i}{p_i}\right)x} H_{0,N}^{N,0} \left[\begin{matrix} - \\ \left\{\left(\frac{d_{i_1}}{p_{i_1}},\frac{1}{p_{i_1}} \right)\right\}_{i_1=1}^{N}\end{matrix} \bigg| \prod_{i_2=1}^{N} \left(\frac{x}{a_{i_2}}\right)\right]\Bigg)
	\label{eq:pdf_h_egg_1}
\end{eqnarray} 
To develop the PDF for the EW channel, we use the Newton's generalized binomial theorem $(1+z)^p=\sum_{j=0}^{\infty}\frac{\Gamma(p+1)z^j}{j!\Gamma(p-j+1)}$ for the term $[1- (\exp(-x/ \eta))^\beta]^{\alpha-1}$ in \eqref{eq:pdf_ew} to get
\begin{eqnarray}
	&f_{h_i}(x)=\frac{\alpha^E_i\beta^E_i\Gamma(\alpha^E_i)}{\eta^E_i}\sum_{j=0}^{\infty}  \frac{(-1)^j}{j!\Gamma(\alpha^E_i-j)}	\bigg( \frac{x}{\eta^E_i}\bigg)^{({\beta^E_i}-1)}\nonumber\\&\exp\bigg(-(j+1)\big(\frac{x}{\eta^E_i}\big)^{\beta^E_i}\bigg)
	\label{eq:pdf_ew_series}	
\end{eqnarray} 
Substituting \eqref{eq:pdf_ew_series} in \eqref{eq:nth_order_moment_of_hi} and applying the identity $\int_{0}^{\infty} t^a \exp(-bt^c)dt=\frac{\Gamma\left(\frac{1+a}{c}\right)}{c b^{\frac{1+a}{c}}}$ \cite[pp. $347$, eq. $3.381.10$]{Zwillinger2014}, we get the $n$-th order moment of $h_{\rm EW}$ as:
\begin{eqnarray}
	&\mathbb{E}(h_{\rm EW}^n)=\prod_{i=1}^{N} \Gamma(\alpha_i+1)\sum_{j=0}^{\infty}\frac{(-1)^j}{j!\Gamma(\alpha_i-j)(j+1)} \nonumber\\& \Gamma\left(1+\frac{n}{\beta_i}\right) \left(\frac{(j+1)^{\frac{1}{\beta_i}}}{\eta_i}\right)
	\label{eq:nth_order_moment_of_h_ew}
\end{eqnarray}
Using \eqref{eq:nth_order_moment_of_h_ew} in \eqref{eq:Nth_order_with_Inverse_Mellin} and applying the definition of Fox H-function to get the PDF of $h_{\rm EW}$ for the EW turbulence:
\begin{eqnarray}
	&f_{h_{\rm EW}}(x)=\prod_{i=1}^{N} \Gamma(\alpha_i+1)\sum_{j=0}^{\infty}\frac{(-1)^j}{j!\Gamma(\alpha_i-j)(j+1)}\nonumber \\& \frac{1}{x}H_{0,N}^{N,0} \left[\begin{matrix} - \\\{(1,\frac{1}{\beta_{i_1}})\}_{i_1=1}^{N} \end{matrix} \bigg|\prod_{i_2=1}^{N} \frac{(j+1)^{\frac{1}{\beta_{i_2}}}x}{\eta_{i_2}}\right]
	\label{eq:pdf_N_EW}
\end{eqnarray} 	

Finally,  we find the PDF of $N$ cascaded $\Gamma\Gamma$ channel $h_{\rm GG}=\prod_{i=1}^{N}h_{i}$, where $h_i$, $i=1,2, \cdots, N$ denote  i.ni.d GG random variables. Note that \cite{Vahid2018} used the method of induction to derive the PDF of cascaded channel for GG oceanic turbulence.
Converting the $v$th-order modified  Bessel's function $K_v(\cdot)$ into  Meijer G function equivalent,  we represent 	\eqref{eq:pdf_gamma_gamma} as
\begin{flalign}
	f_{h_i}(x)=\frac{(\alpha^G\beta^G)^{(\alpha^G+\beta^G)/2}}{\Gamma({\alpha^G})\Gamma({\beta^G})}x^{\frac{\alpha^G+\beta^G}{2}-1} \nonumber \\ {G^{2,0}_{0,2}\Bigg(\alpha\beta x\bigg|\begin{matrix}-\\\ \frac{\alpha^G-\beta^G}{2},\frac{\beta^G-\alpha^G}{2}
		\end{matrix}\Bigg)}
	\label{eq:pdf_gamma_gamma_MG}
\end{flalign}
Substituting \eqref{eq:pdf_gamma_gamma_MG} in \eqref{eq:nth_order_moment_of_hi} and applying the identity \cite[eq. $07.34.21.0009.01$]{Wolfram}, we get the $n$-th order moment of $h_{\Gamma\Gamma}$ as: 
\begin{equation}
	\mathbb{E}(h_{\Gamma\Gamma}^n)=\prod_{i=1}^{N} \frac{(\alpha_i\beta_i)^{-n}}{\Gamma(\alpha_i)\Gamma(\beta_i)}\Gamma(\alpha_i+n)\Gamma(\beta_i+n)
	\label{eq:nth_order_moment_of_h_gg}
\end{equation}

We use \eqref{eq:nth_order_moment_of_h_gg} in \eqref{eq:Nth_order_with_Inverse_Mellin} and apply the definition of Fox H-function to get the PDF of $h_{\Gamma\Gamma}$ as  
\begin{flalign}
	&f_{h_{\Gamma\Gamma}}(x)=\prod_{i=1}^{N} \frac{1}{\Gamma(\alpha_i)\Gamma(\beta_i)}\frac{1}{x}\nonumber \\& H_{0,2N}^{2N,0} \left[\begin{matrix} - \\\{(\alpha_{i_1},1 )\}_{i_1=1}^{N},\{(\beta_{i_1},1 )\}_{i_1=1}^{N} \end{matrix} \bigg|\prod_{i_2=1}^{N} \alpha_{i_2}\beta_{i_2}x\right]
	\label{eq:pdf_N_gamma_gamma}
\end{flalign}

Capitalizing \eqref{eq:pdf_N_gen_gamma}, \eqref{eq:pdf_h_egg_1}, 	\eqref{eq:pdf_N_EW}, \eqref{eq:pdf_N_gamma_gamma}, we get  an expression for the unified PDF in \eqref{eq:unified_uw_pdf} with parameters as depicted Table \ref{table1}, which concludes the proof for Theorem 1. 

\section*{Appendix B: Proof of Lemma 1}
Using the product distribution \cite{Papoulis2001}, the PDF of the combined channel $h=h_ch_p$ can be expressed as 
\begin{eqnarray}
	f_{h}(h)=\int_{0}^{A_0}\frac{1}{h_p}f_{h_p}(h_p)f_{h_c}\left(\frac{h}{h_p} \right)dh_p
	\label{eq:combined_hc_hp}
\end{eqnarray}
We substitute \eqref{eq:pdf_hp} and \eqref{eq:unified_uw_pdf} in \eqref{eq:combined_hc_hp}, solve the inner integral $\int_{0}^{A_U}h_p^{\rho_U^2+n-1}\diff h_P=\frac{A_U^{\rho_U^2+n}}{(\rho_U^2+n)}=\frac{A_0^{\rho_U^2+n}\Gamma(\rho_U^2+n)}{\Gamma(1+\rho_U^2+n)}$  and apply the definition of Fox H-function \cite{mathai2009} to get
\begin{eqnarray}
	&f_{h}(h)=\prod_{i=1}^{N}\rho_U^2\mathcal{A}_i \sum_{j=0}^{\mathcal{P}}\mathcal{B}_{i,j}\nonumber\\&\frac{1}{h}H_{1,q+1}^{m+1,0} \left[\begin{matrix} (1+\rho_U^2) \\\{\mathcal{D}_{i_1,j}\}_{i_1=1}^N, (\rho_U^2,1) \end{matrix} \bigg|\prod_{i_2=1}^{N} \frac{\mathcal{E}_{i_2,j} h}{A_U} \right]
	\label{eq:pdf_hc_hp2} 
\end{eqnarray}
Thus, we use  the transformation of random variable $\gamma= {\gamma_U} h^2$ to get the PDF of SNR in \eqref{eq:pdf_unified_hp}. To find the CDF of SNR under the combined channel, we use \eqref{eq:pdf_unified_hp} in $ F_{\gamma}(\gamma)=\int_0^{\gamma}f(\gamma)d\gamma$
and apply the definition of Fox H-function with inner integral $\int_{0}^{\gamma}\gamma^{-\frac{n}{2}-1} d\gamma=\frac{\gamma^{-\frac{n}{2}}}{-\frac{n}{2}}=\frac{2\gamma^{-\frac{n}{2}}\Gamma(-n)}{\Gamma(1-n)}$ to get the CDF of the SNR in \eqref{eq:cdf_unified_hp}, which concludes the proof of Lemma \ref{theorem_unified_hp}.	
\section*{Appendix C: Proof of Lemma 2}
Using \eqref{eq:pdf_fpt_twoc} and \eqref{eq:pdf_unified_hp} in 	\eqref{eq:snr_pdf_eqn_af_gen} results the PDF of the mixed link as
\begin{align} \label{eq:snr_pdf_eqn_af_1}
	&f_{\gamma}(\gamma) = \frac{z^k\rho_T^2A^{\rm{mg}}}{4 \gamma}  \sum_{m_1=1}^{\beta^M}b^M_{m_1}  \bigg(\frac{1}{2\pi \J}\bigg)^{2} \int_{\mathcal{L}_{1}}\int_{\mathcal{L}_{2}} \prod_{i=1}^{N}\frac{\rho_U^2\mathcal{A}_i}{2} \nonumber\\&\sum_{j=0}^{\mathcal{P}}\mathcal{B}_{i,j} \bigg(\frac{\alpha^M\beta^M \sqrt{\gamma}}{(g^M\beta^M+\Omega^M)A_T\sqrt{\bar{\gamma}_T}}\bigg)^{n_{1}}\nonumber \\ & \frac{\Gamma(\rho_T^{2}-n_{1})\Gamma(\alpha^{M}-n_{1})\Gamma(m_1-n_{1})[\Gamma(z-n_{1})]^k_1}{\Gamma(\rho_T^{2}+1-n_{1})[\Gamma(z+1-n_{1})]^k_1} \nonumber \\ &\bigg(\frac{\mathcal{E}_{i_2,j}}{A_U} \sqrt{\frac{1}{\bar{\gamma}_U}}\bigg)^{-n_{2}}  \frac{\prod_{j=1}^{m} [\Gamma(\mathcal{D}_{i,j})]^{N}_{i=1}\Gamma(\rho_{U}^{2}+n_{2})}{\Gamma(\rho_{U}^{2}+1+n_{2})}  \nonumber \\ &\bigg(\int_{0}^{\infty} x^{-1-\frac{n_{2}}{2}} \bigg(\frac{x+C}{x}\bigg)^{\frac{n_{1}}{2}} \diff x\bigg) \diff n_{1} \diff n_{2}  
\end{align}

We use \cite[(3.194.3)]{Zwillinger2014}  and \cite[(8.384.1)]{Zwillinger2014}  to solve the inner integral in terms of Gamma functions:
\begin{equation} \label{eq:snr_pdf_eqn_af_2}
	\int_{0}^{\infty} x^{-1-\frac{n_{2}}{2}} \bigg(\frac{x+C}{x}\bigg)^{\frac{n_{1}}{2}} \diff x = \frac{C^{-\frac{n_{2}}{2}}\Gamma(\frac{n_{2}}{2})\Gamma(-\frac{n_{1}}{2}-\frac{n_{2}}{2})}{\Gamma(-\frac{n_{1}}{2})}
\end{equation}
Substitute \eqref{eq:snr_pdf_eqn_af_2} in \eqref{eq:snr_pdf_eqn_af_1}, we get
\begin{align} \label{eq:snr_pdf_eqn_af_3}
	&f_{\gamma}(\gamma) = \frac{z^k\rho_T^2A^{\rm{mg}}}{4 \gamma}  \sum_{m_1=1}^{\beta^M}b^M_{m_1}  \bigg(\frac{1}{2\pi \J}\bigg)^{2} \int_{\mathcal{L}_{1}}\int_{\mathcal{L}_{2}} \prod_{i=1}^{N}\frac{\rho_U^2\mathcal{A}_i}{2} \nonumber\\&\sum_{j=0}^{\mathcal{P}}\mathcal{B}_{i,j} \bigg(\frac{(g^M\beta^M+\Omega^M)A_T\sqrt{\bar{\gamma}_T}}{\alpha^M\beta^M \sqrt{\gamma}}\bigg)^{-n_{1}}\nonumber \\ & \frac{\Gamma(\rho_T^{2}-n_{1})\Gamma(\alpha^{M}-n_{1})\Gamma(m_1-n_{1})[\Gamma(z-n_{1})]^k_1}{\Gamma(\rho_T^{2}+1-n_{1})[\Gamma(z+1-n_{1})]^k_1\Gamma(-\frac{n_{1}}{2})} \nonumber \\ &\bigg(\frac{\mathcal{E}_{i_2,j}}{A_U} \sqrt{\frac{1}{\bar{\gamma}_U}}\bigg)^{-n_{2}}  \frac{\prod_{j=1}^{m} [\Gamma(\mathcal{D}_{i,j})]^{N}_{i=1}\Gamma(\rho_{U}^{2}+n_{2})\Gamma(\frac{n_{2}}{2})}{\Gamma(\rho_{U}^{2}+1+n_{2})}   \nonumber\\&\Gamma(-\frac{n_{1}}{2}-\frac{n_{2}}{2}) \diff n_{1} \diff n_{2} 
\end{align}
Thus, we apply the definition of bivariate Fox-H function \cite{Mittal_1972} to get \eqref{eq:snr_pdf_eqn_af}.

Similarly,  we use  \eqref{eq:snr_pdf_eqn_af} in $F(\gamma)=\int_0^\gamma f(\gamma)$ to derive the CDF to get  
\begin{align} \label{eq:snr_cdf_eqn_af_1}
	&F_{\gamma}(\gamma) = \frac{z^k\rho_T^2A^{\rm{mg}}}{4 }  \sum_{m_1=1}^{\beta^M}b^M_{m_1}  \bigg(\frac{1}{2\pi \J}\bigg)^{2} \int_{\mathcal{L}_{1}}\int_{\mathcal{L}_{2}} \prod_{i=1}^{N}\frac{\rho_U^2\mathcal{A}_i}{2} \nonumber\\&\sum_{j=0}^{\mathcal{P}}\mathcal{B}_{i,j} \bigg(\frac{(g^M\beta^M+\Omega^M)A_T\sqrt{\bar{\gamma}_T}}{\alpha^M\beta^M \sqrt{\gamma}}\bigg)^{-n_{1}}\nonumber \\ & \frac{\Gamma(\rho_T^{2}-n_{1})\Gamma(\alpha^{M}-n_{1})\Gamma(m_1-n_{1})[\Gamma(z-n_{1})]^k_1}{\Gamma(\rho_T^{2}+1-n_{1})[\Gamma(z+1-n_{1})]^k_1\Gamma(-\frac{n_{1}}{2})} \nonumber \\ &\bigg(\frac{\mathcal{E}_{i_2,j}}{A_U} \sqrt{\frac{1}{\bar{\gamma}_U}}\bigg)^{-n_{2}}  \frac{\prod_{j=1}^{m} [\Gamma(\mathcal{D}_{i,j})]^{N}_{i=1}\Gamma(\rho_{U}^{2}+n_{2})\Gamma(\frac{n_{2}}{2})}{\Gamma(\rho_{U}^{2}+1+n_{2})}   \nonumber\\&\Gamma\left(-\frac{n_{1}}{2}-\frac{n_{2}}{2}\right) \bigg(\int_{0}^{\gamma} \gamma^{-\frac{n_{1}}{2}-1}\diff \gamma\bigg) \diff n_{1} \diff n_{2} 
\end{align}
The inner integral can be solved as
\begin{flalign}
	\label{eq:inner}
\int_{0}^{\gamma} \gamma^{-\frac{n_{1}}{2}-1}\diff \gamma=-\frac{\gamma^{-\frac{n_{1}}{2}}}{\frac{n_{1}}{2}}=\gamma^{-\frac{n_{1}}{2}}\frac{\Gamma(-\frac{n_{1}}{2})}{\Gamma(1-\frac{n_{1}}{2})}
\end{flalign} 
Using \eqref{eq:inner} in \eqref{eq:snr_cdf_eqn_af_1}, we get \eqref{eq:snr_cdf_eqn_af}, which completes the proof of Lemma \ref{lemma:pdfcdf_af}.

\bibliographystyle{ieeetran}
\bibliography{bib_file}

    \end{document}